\documentclass{article}

\usepackage{arxiv}

\usepackage[utf8]{inputenc} 
\usepackage[T1]{fontenc}    
\usepackage{hyperref}       
\usepackage{url}            
\usepackage{booktabs}       
\usepackage{amsfonts}       
\usepackage{nicefrac}       
\usepackage{microtype}      
\usepackage{lipsum}
\usepackage{graphicx}
\graphicspath{ {./images/} }
\usepackage{todonotes}
\usepackage[acronym]{glossaries}
\usepackage{caption}
\usepackage{subcaption}
\usepackage{amsmath}
\usepackage{amsfonts}
\usepackage{amssymb}
\usepackage{float}
\usepackage{alphabeta}
\usepackage{CJKutf8}
\usepackage{graphicx}
\usepackage{enumerate}
\usepackage{xcolor}
\usepackage{natbib}

\usepackage{prodint}
\usepackage[printonlyused,withpage]{acronym}
\usepackage[titletoc]{appendix}
\usepackage{hyperref}
\usepackage{chemfig}
\usepackage{longtable}
\usepackage{emptypage}
\usepackage{lscape}
\usepackage{longtable}
\usepackage{color, colortbl}

\makeglossaries
\newacronym{ann}{ANN}{Artificial Neural Network}
\newacronym{auroc}{AUROC}{Area Under the Receiver Operating Characteristic}
\newacronym{auprc}{AUPRC}{Area Under the Precision-Recall Curve}
\newacronym{mcc}{MCC}{Matthews Correlation Coefficient}
\newacronym{1dcnn}{1-D CNN}{1-D Convolutional Neural Network}
\newacronym{2dcnn}{2-D CNN}{2-D Convolutional Neural Network}
\newacronym{1ddnn}{1-D DNN}{1-D Deconvolutional Neural Network}
\newacronym{dnn}{DNN}{Deconvolutional Neural Network}
\newacronym{cnn}{CNN}{Convolutional Neural Network}
\newacronym{dwt}{DWT}{Discrete Wavelet Transform}
\newacronym{mlp}{MLP}{Multi-Layer Perceptron}
\newacronym{lstm}{LSTM}{Long Short-Term Memory}
\newacronym{1dcnnlstm}{1D-CNN-LSTM}{1-D Convolutional Neural Network with Long Short-Term Memory}
\newacronym{auc}{AUCROC}{Area Under the Receiver Operating Characteristic}
\newacronym{sem}{SEM}{Standard Error Mean}
\newacronym{dl}{DL}{Deep Learning}
\newacronym{ks}{KS}{K-Shape}
\newacronym{pca}{PCA}{Principal Component Analysis}
\newacronym{lr}{LR}{Logistic Regression}
\newacronym{pme}{PME}{Principle of Maximum Entropy}
\newacronym{raf}{RAF}{Rational Activation Function}
\newacronym{relu}{ReLU}{Rectified Linear Unit}
\newacronym{roar}{ROAR}{Remove and Retain}
\newacronym{kar}{KAR}{Keep and Retain}
\newacronym{xai}{XAI}{Explainable Artificial Intelligence}
\newacronym{sm}{SM}{Saliency Map}
\newacronym{kpca}{KPCA}{Kernel Principal Component Analysis}
\newacronym{knn}{KNN}{K-Nearest Neighbors}
\newacronym{smote}{SMOTE}{Synthetic Minority Over-sampling Technique}
\newacronym{enn}{ENN}{Edited Nearest Neighbors}
\newacronym{smoteenn}{SMOTEENN}{Synthetic Minority Over-sampling Technique and Edited Nearest Neighbors}
\newacronym{3mcs}{3MCS}{3-metric combined score}
\newacronym{cssm}{CSSM}{Chi-Squared Saliency Map}
\newacronym{lrt}{LRT}{ Likelihood Ratio Test}

\newacronym{ehr}{EHR}{Electronic Health Record}
\newacronym{tsi}{TSI}{Time Series Instance}
\newacronym{icu}{ICU}{Intensive Care Unit}
\newacronym{ecg}{ECG}{Electrocardiogram}
\newacronym{nsr}{NSR}{Normal Sinus Rythm}
\newacronym{af}{AF}{Atrial Fibrillation}
\newacronym{mi}{MI}{Myocardial Infarction}
\newacronym{sbr}{SBR}{Sinus Bradycardia Rythm}

\title{Constructing Interpretable Prediction Models with 1D DNNs: An Example in Irregular ECG Classification}

\author{
 Giacomo Lancia \\
  Department of Basic and Applied Sciences for Engineering (SBAI)\\
  University of Rome "La Sapienza"\\
  Via Antonio Scarpa, 4. 00161, Roma.\\
  \texttt{giacomo.lancia@uniroma1.it} \\
   \And
    Cristian Spitoni \\
  Department of Mathematics\\
  Utrecht University\\
  Budapestlaan, 6. 3584 CD, Utrecht.\\
  \texttt{c.spitoni@uu.nl} \\
}

\begin{document}
\maketitle
\begin{abstract}
This manuscript proposes a novel methodology aiming at developing an interpretable prediction model for irregular \gls{ecg} classification, by using features extracted by a \gls{1ddnn}. Given the increasing prevalence of cardiovascular disease, there is indeed a growing demand for models that provide transparent and clinically relevant predictions, which are essential for advancing the development of automated diagnostic tools.
The features extracted by the \gls{1ddnn} are then included in a simple \gls{lr} model, in order to predict abnormal \gls{ecg} patterns.  Our analysis shows in a nutshell that the features are consistent with the clinical knowledge, and provide an interpretable and reliable classification of conditions like \gls{af}, \gls{mi}, and \gls{sbr}.
Moreover, our findings demonstrate that the simple \gls{lr}  has similar predictive accuracy to one of the more complex models, such as a\gls{1dcnn}, providing a concrete example of how to efficiently integrate \gls{xai} methodologies with traditional regression models.

\end{abstract}

\section{Introduction}
When considering \gls{dl}-based applications in medicine, the interpretation of results represents a primary aspect of assessing the quality and safety of clinical predictions \cite[]{stiglic2020interpretability}.
On the other hand, as massive databases of complex high-dimensional data have become more available, the demand for accurate computational models capable of capturing salient information from data has become predominant \cite[]{norori2021addressing, pereira2021sharing, jiang2017artificial}.
In recent years, \gls{ann}-based methods have proven to be a promising approach for achieving highly accurate predictions across many areas of medical research \cite[]{elkhader2022artificial, radakovich2020artificial, luchini2022artificial, winchester2023artificial, van2023successfully}, providing improvements to more traditional approaches \cite[]{lauritsen2020explainable, zihni2020opening, keyl2022patient, chan2022explainable, lancia2023learning}.
Unfortunately, due to the intrinsic \emph{black box} nature of ANN methods, the causal relation between covariates and prediction cannot be easily assessed. Hence, integrating the \gls{ann} potentially into an interpretable framework becomes therefore necessary in order to guarantee the reliability and the transparency of the models \cite[]{rudin2019stop}. 

\gls{sm}-based methods, and more in general \gls{xai} algorithms, have been recently proposed as a novel methodology to explain the activity of plenty of \gls{ann} architectures (\cite{chopra2022evaluation, kumar2022overview}).
The basic idea of \gls{sm} is to visualize the areas of the input data where the \gls{ann} brings its focus.
\gls{sm}s therefore provide an estimate of the degree of importance according to the \gls{ann} across the entire input data domain.
In the framework of feed-forward \gls{ann}s, the construction of a \gls{sm}-based explanatory method is then achieved through a supplementary model that performs a sequence of algorithmic rules; no additional training phase is required. 
These algorithmic rules are usually devised to exploit the peculiarities of the feed-forward \gls{ann} network's forward propagation rule, aiming to understand the backward connection between outputs and inputs. 
For example, the Vanilla Gradient \cite[]{simonyan2013deep} represented one pioneeristic attempt to determine the change of outputs with respect to inputs by exploiting the backpropagation updating rule of feed-forward \gls{ann}. 
Depending on the architecture of the \gls{ann} to be explained, various algorithms, whether specific or general, have been proposed.
Algorithms like CAM \cite[]{zhou2016learning}, Layer-Wise Relevance Propagation \cite[]{montavon2019layer} and \gls{dnn} \cite{Zeiler2014} represent a few valid examples among the most popular backpropagation-based methods for explaining the pattern recognition activity of \gls{cnn} models.
Regarding the specific case of \gls{cnn}s-based models built on the attention mechanism, a few solutions to explain these models have been recently proposed in \cite{hasanpour2022unboxing, gkartzonika2022learning}.
With a particular focus on medical imaging, the utilization of \gls{sm}-based applications in numerous sub-fields has also gained considerable attention \cite[]{hossain2023explainable, fontes2024application}. 

In light of this, we propose in this work a novel and interpretable \gls{ann}-based methodology with the ultimate scope of classifying abnormalities in \gls{ecg} records.
According to the Heart Disease and Stroke Statistics report, the rapid aging of the global population has made cardiovascular disease the foremost cause of death \cite[]{tsao2022heart}.
Early detection of arrhythmia can help identify potential risk factors for heart attacks \cite{akter2024identification}.
Similarly, the increasing prevalence of myocardial infarction underscores the critical need for accurate and prompt diagnosis to improve patient treatment \cite[]{ abagaro2024automated}.
The demand for real-time interpretable models for irregular \gls{ecg} detection is becoming even more predominant to ensure timely and efficient predictions that are also explainable and understandable during patient monitoring \cite[]{liu2024lightweight}.
As well, intuitive interfaces and rapid visualizations of interpretable predictions provide useful instrumentation to help clinicians diagnose \gls{ecg} irregularities effectively \cite[]{nguyen2024bibliometric, sharma2022explainable}.
However, standardized benchmarks and metrics are necessary to evaluate and compare the effectiveness of interpretability methods applied to \gls{ann}, guiding their deployment in clinical settings effectively \cite{chopra2022evaluation}.

Therefore, our research aims to explore how one \gls{ann} processes a large volume of sequential \gls{ecg} data in order to leverage its ability to make robust, interpretable, and accurate predictions for \gls{ecg} data. 
Specifically, we seek to establish if the decision-making process of \gls{ann} activity aligns with existing clinical knowledge for \gls{ecg} prediction.
Equally important, we aim to find out if this investigation can provide new insights into modelling irregular \gls{ecg} patterns. 
Our goal is to make the modelling process easier and more interpretable while ensuring at the same time accuracy and clinical reliability. 
Although the classification of \gls{ecg} irregularities has garnered significant attention in the last decade, most proposed methodologies have focused either on efficient \emph{ad-hoc} clinical-based data pre-processing (e.g., \cite{sharma2022explainable, sharma2024intelligent}) or on utilizing powerful yet difficult-to-interpret \gls{ann}s methodologies to finalize predictions (e.g., \cite{eleyan2024electrocardiogram, katal2023deep}.
In the former approach, the type of pre-processing used can impact the study's outcome. It may require significant effort to develop a pre-processing stage that captures the essential features needed for accurate predictions in a broad setting.
In the latter approach, as mentioned, predictions may not accurately reflect clinical reality and can be difficult to interpret.
It is important to note that \gls{xai}-based models have emerged as valuable tools for understanding the potential causal relationships between input \gls{ecg} data and their classification (e.g., \cite{ojha2024exploring, han2023automated}).
Anyway, in both general and specific settings, \gls{sm}s derived from \gls{xai} methods mainly help visualize where the \gls{ann} focused; they do not reveal the causal connections between the available data and the phenomenon intended to be modelled.
As a result, interpretations based solely on \gls{xai} methods might hide the risk of drawing incorrect conclusions. 

Building on the approach proposed in \cite{lancia2024two} for making \gls{1dcnn}-based predictions interpretable, we propose an automated methodology to capture and extract the essential features of \gls{1dcnn} pattern recognition.
The ultimate scope is to utilize these extracted features to formulate new interpretable predictions to classify irregular \gls{ecg}.
To ensure the interpretability of these predictions, we then use a \gls{lr} model.
Due to its simple structure, such a model is often regarded as easy to interpret.
Consequently, using a feature extraction method based on the analysis of \gls{sm}s derived from \gls{1ddnn} activity enhances the predictive power of the \gls{lr}.
classifying \gls{ecg} data. 
In addition, a \emph{post-hoc} inspection of the \gls{lr} model is then accomplished to determine whether the formulated predictions align with the clinical assessments expected from clinicians while identifying \gls{ecg} abnormalities.

This paper is organized as follows: Section \ref{sec: Datasets} presents a detailed overview of the \gls{ecg} dataset, which was acquired from the open-access source Physionet \cite{goldberger2000physiobank}. In Section \ref{sec: ECG Pre-processing}, we introduce the data, describe the pre-processing steps, and outline the classification problems we aim to address.
Section  \ref{sec:model_fit} focuses on resolving the irregular \gls{ecg} problem using \gls{1dcnn}s. Following this, in Section \ref{sec: 1dcnn_inversion}, we propose a methodology for efficiently reconstructing \gls{1dcnn} inputs through a \gls{1ddnn}, considering reconstructions derived from the deepest layers of the \gls{1dcnn}.
From a methodological perspective, we will demonstrate that employing semi-orthogonal convolutional kernels in the \gls{1dcnn} is crucial for enhancing the performance of the \gls{1ddnn} and achieving more precise reconstructions. This novel approach significantly improves the power of feature extraction derived from the \gls{1ddnn}.
Section \ref{sec: 1-D DNN-based saliency maps} discusses the construction of saliency maps from the \gls{1ddnn} reconstructions. 
In Section \ref{sec: feature extraction}, we will focus on extracting novel, easy-to-understand, and efficient features of irregular \gls{ecg} based on the explainability of the \gls{1dcnn}.
We then propose an interpretable model, utilizing the extracted features from Section \ref{sec: feature extraction}, in Section \ref{sec: interpretable predictions}. 
The performance and explainability of the \gls{1dcnn} are evaluated in Section \ref{sec:results}, and we examine the interpretable predictions in Section \ref{sec: Inspecting_Interpretable_Predictions}.
Finally, a brief discussion is included in Section \ref{sec:discussion}.

\section{Data}\label{sec:data}

We utilized data available from the free-access research database for 12-lead electrocardiogram (ECG) signals, created by Chapman University, Shaoxing People’s Hospital, and Ningbo First Hospital \cite[]{zheng2022large}.
This dataset was originally presented in \cite{zheng2020optimal}; it is freely available from the Physionet \cite{goldberger2000physiobank}. 
The database comprehends a vast collection of 12-lead \gls{ecg} records from 45152 patients at a 500 Hz sampling rate, labeled by professionals.
Among the data, one may find different \gls{ecg}s of various kinds of 10-second \gls{ecg} records of irregular and regular heartbeats.
The final aim of this database is to support the development and refinement of machine learning and statistical tools for automated diagnosis of irregular cardiovascular rhythms.

\begin{figure}
    \centering
    \includegraphics[height=7cm]{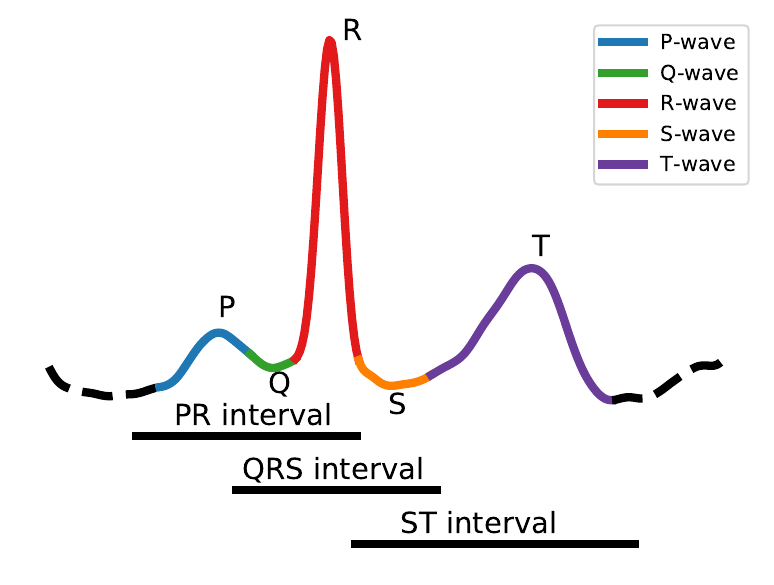}
    \caption{An example of a normal heart's beat. 
    The P-wave is depicted in blue, while the T-wave is in purple. 
    The QRS complex is also present with the Q-wave in red, the R-peak in green, and the S-wave in red.
    Intervals are shown underneath the corresponding waves}
    \label{fig: Schematic_HeartBeat}
\end{figure}

\gls{ecg}s are graphical representations of the heart's electrical activity over time, showing the processes of cardiac muscle depolarization and repolarization during each heartbeat. 
Typically, a normal heart's beat includes the P-wave (atrial depolarization), the QRS complex (ventricular depolarization), and the T-wave (ventricular repolarization), along with intervals such as PR, ST, and QT; see Figure \ref{fig: Schematic_HeartBeat}. 

Each \gls{ecg} was acquired over a period of 10 seconds, and it was accomplished by subjecting each patient to a 12-lead resting \gls{ecg} test.
A licensed physician labeled the acquired \gls{ecg} and the cardiac conditions, with a secondary validation by another physician.
A senior physician resolved possible disagreements occurring during the validation process.
Together with the \gls{ecg} signals, baseline covariates such as age and gender were also acquired.

\subsection{Datasets}\label{sec: Datasets}


The work primarily focused on constructing \gls{ann}-based interpretable prediction models, each focused on a specific cardiac pathology of interest.
To achieve this, we utilized all the available records to construct three different datasets.
Each dataset comprised copies of the complete collection of \gls{ecg}s, allowing us to identify pathological conditions such as \gls{af}, \gls{mi}, and \gls{sbr} in isolation.
For each dataset, we labeled the \gls{ecg}s associated with the pathology of interest as cases, while the remaining records served as controls. 
For instance, in the binary classification task aimed at predicting \gls{af}, all \gls{ecg}s identified as \gls{af} were categorized as cases, while those labeled as \gls{mi} and \gls{sbr} were assigned to the control group.
The overall cohort of patients consisted exclusively of adults aged 18 and older; the median age is 63 with an interquartile range of 21. 
Minimal and maximal ages were 18 and 89.
The population consisted of 56\% of males and 44\% of females.

The sub-population experiencing \gls{mi} represented a large minority of the cohort (circa 0.02\%), while other sub-populations, such as those experiencing \gls{af} and \gls{sbr}, are of 3\% and 36\%, respectively.
Subjects with a \gls{nsr}, i.e., a clinical situation where no \gls{ecg} irregularity is present, are 17\% of the population. 
These sub-populations' ages mainly range between 50 and 80, with a net prevalence of male subjects.
A summary statistic is reported in table \ref{tab: Summary_Statistic}.
\begin{table}[h]
    \centering
    \begin{tabular}{|l|c|c|c|c|}
    \hline
      \textbf{Rythm}  & \textbf{Sample size} & \textbf{Age (Median -- IQR)}& \textbf{Male (\%)} & \textbf{Female (\%)} \\ 
    \hline
       All ECGs  & 45152 &  63 -- 21 &  56\% &44\% \\
       Sinus Rhythm &7751& 55 -- 15& 43\%& 57\% \\
       Atrial Fibrillation  &  1780 & 75 -- 15 &  58\%& 62\%\\
       Myocardial Infarction  & 120 & 69 -- 25 & 76\%&  24\%\\
       Sinus Bradycardia Rythm  & 16417 & 60 -- 18 &  63\%& 37\%\\
    \hline
       
    \end{tabular}
    \caption{Summary statistics of Sample size, age, and gender stratified per type of ECG}
    \label{tab: Summary_Statistic}
\end{table}
\begin{figure}
    \centering
    \includegraphics[width= .58\textwidth]{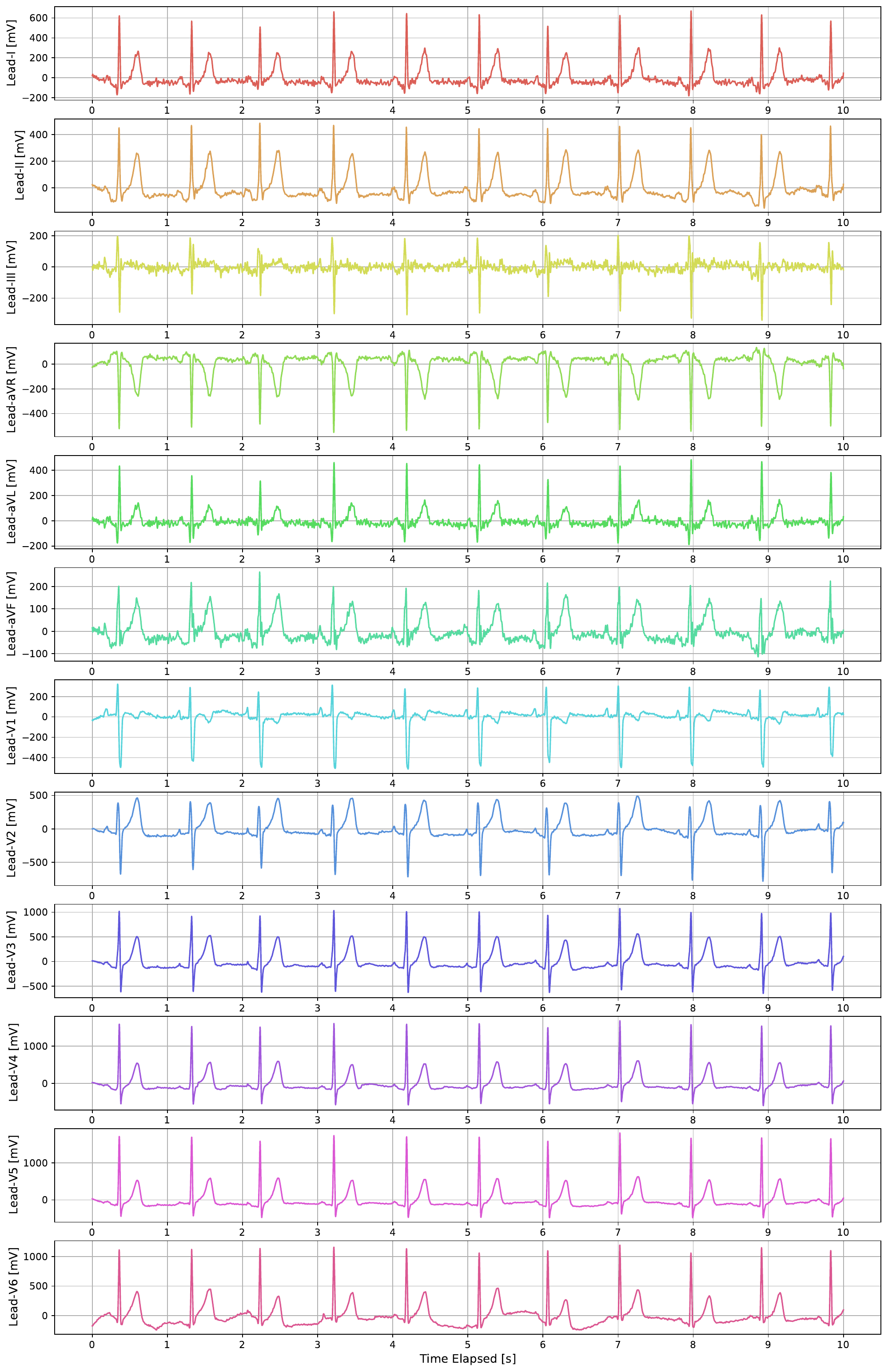}
    \caption{An example of 12-lead ECG for Sinus Rhythm}
    \label{fig: example_Sinus_Rythm}
\end{figure}
\begin{figure}
    \centering
    \includegraphics[width= .58\textwidth]{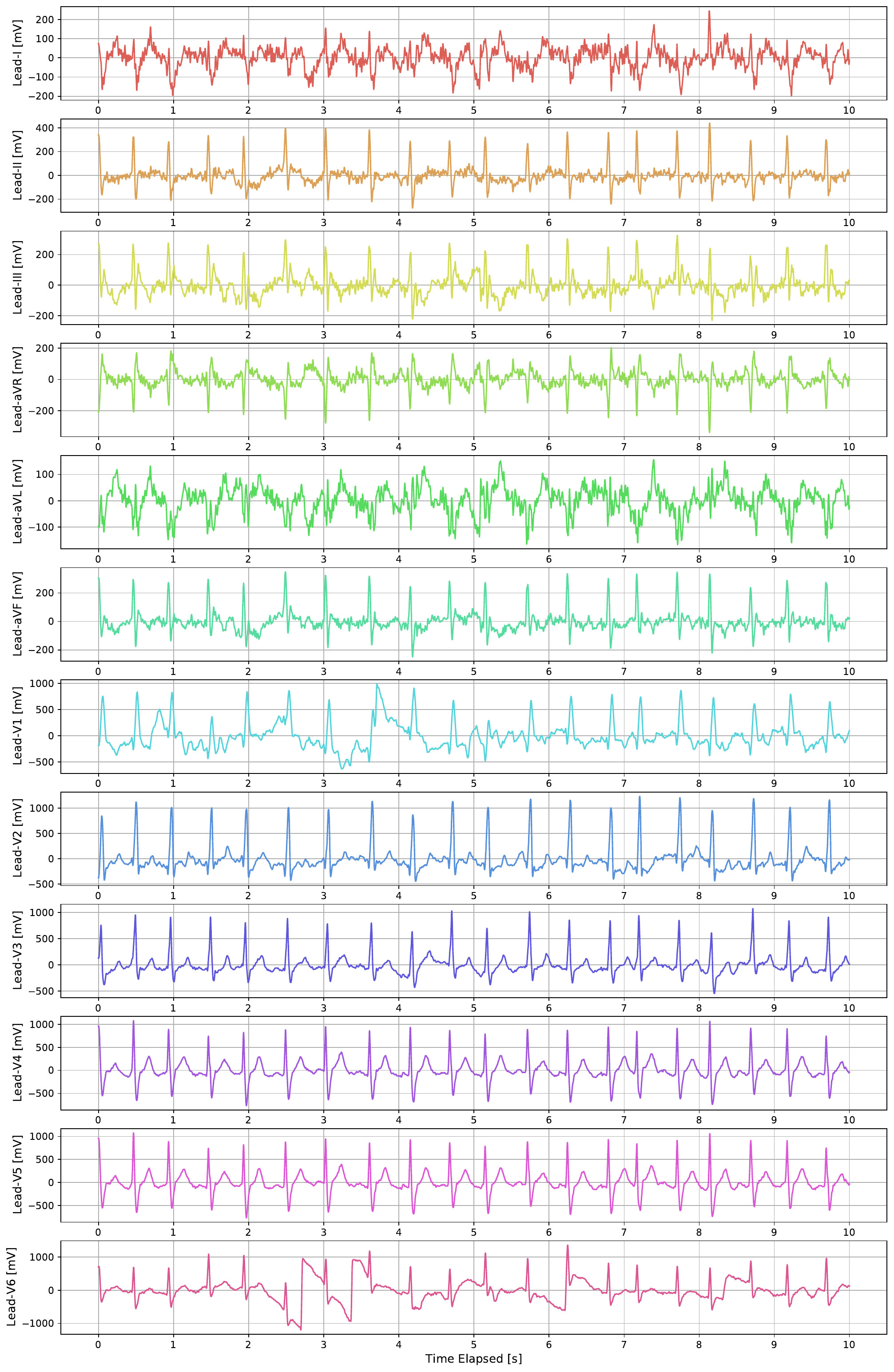}
    \caption{An example of 12-lead ECG for Atrial Fibrillation}
    \label{fig: example_Atrial_Fibrillation}
\end{figure}
\begin{figure}
    \centering
    \includegraphics[width= .58\textwidth]{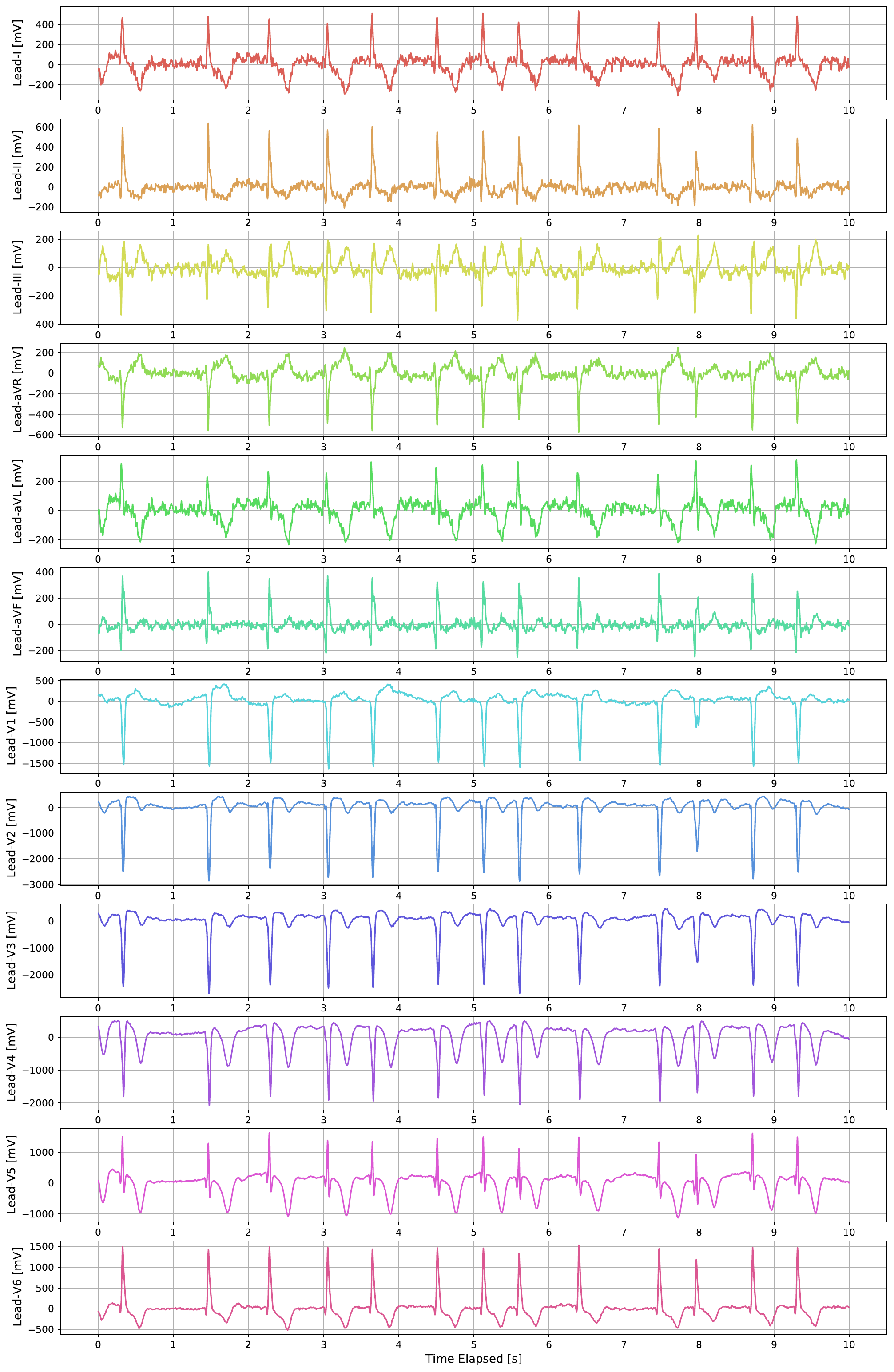}
    \caption{An example of 12-lead ECG for Myocardial Infarction}
    \label{fig: example_myocardial_infarction}
\end{figure}
\begin{figure}
    \centering
    \includegraphics[width= .58\textwidth]{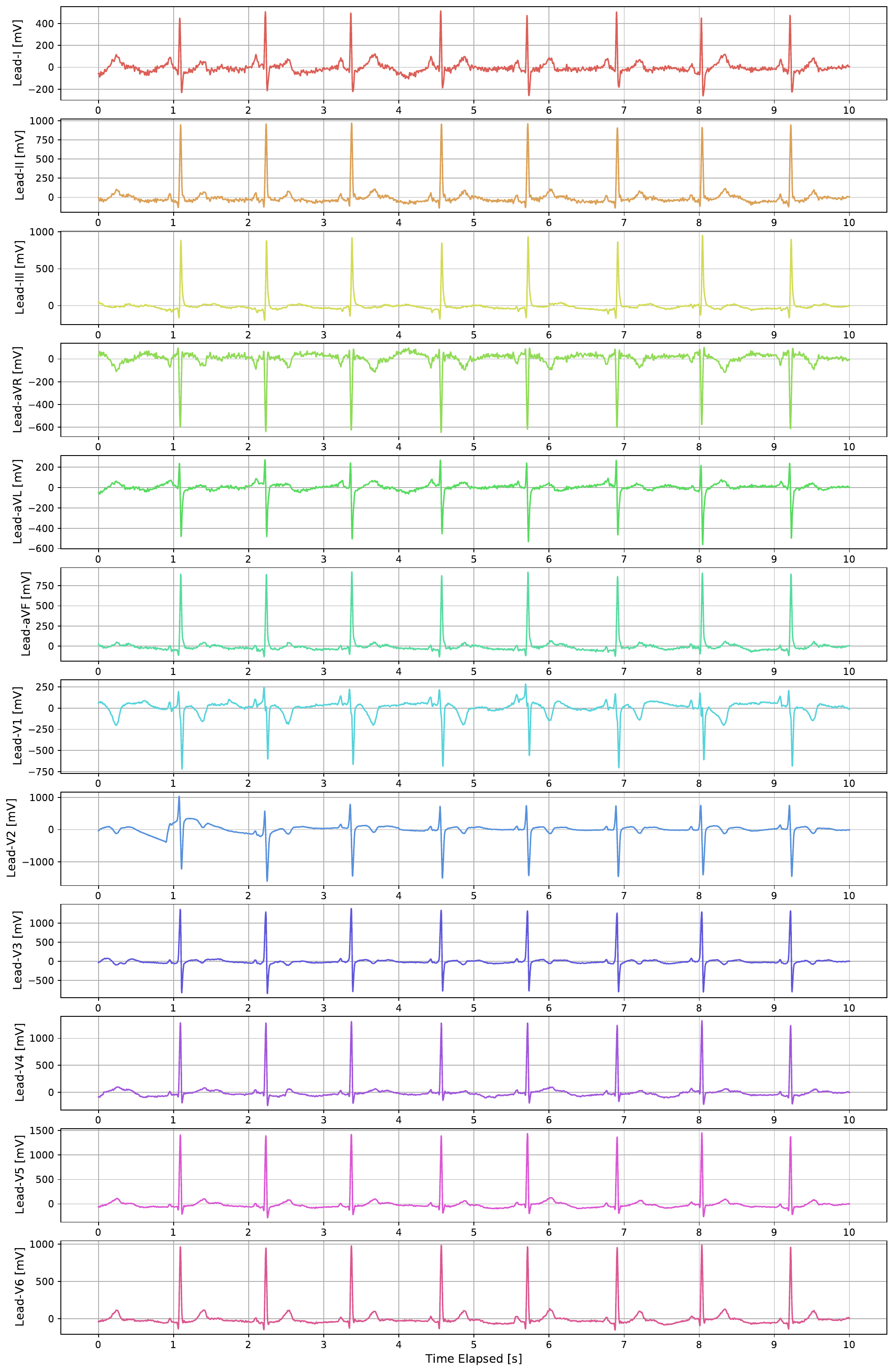}
    \caption{An example of 12-lead ECG for Sinus Bradycardia Rythm}
    \label{fig: example_Sinus_Bradycardia}
\end{figure}

\subsection{ECG Pre-processing}\label{sec: ECG Pre-processing}
The raw \gls{ecg} records, as taken from the acquisition stage, might appear blurred.
A pre-processing stage was therefore needed to tackle the effects of all sources of noise interfering with the heart's electrical response.
Blurring effects are typically due to power line interference, electrode contact noise, motion artifacts, skeletal muscle contraction, baseline wandering, and other random noise occurring during the acquisition.
All these sources of noise might lay on specific spectral bands or be spread along all the spectral domains.
For example, the power line interference occurs at a frequency of 50-60 Hz while the frequency of the baseline wander is less than 0.5 Hz \cite{zheng2020optimal}.

Based on these considerations, we combined the Butterworth filtering method and the Savitzy-Golay method to attempt to remove the desired noise. 
In particular, we first use in sequence two Butterworth filters, a low-pass and a high-pass filter with cutoff at 50 Hz and 0.5 Hz, respectively.
This stage aimed to remove the noisy effects due to power line interference and baseline wander.
To reinforce the elimination of the baseline wander, we then utilized the Savitzy-Golay method.

Examples of pre-processed \gls{ecg}s are shown in Figures \ref{fig: example_Sinus_Rythm}-\ref{fig: example_Sinus_Bradycardia}.
In particular, for \gls{nsr} the P and T waves, together with the QRS complex, are well visible and pointing upwards, while the structures of the lead avF are pointing downwards (see fig. \ref{fig: example_Sinus_Rythm}).
For \gls{af}, the absence of P wave appears evidently, e.g., leads V3-V6, and II (see fig. \ref{fig: example_Atrial_Fibrillation}).
For \gls{mi}, an alteration of the QRS complex is evident from leads V3-V6 and V1-V4 (see fig. \ref{fig: example_myocardial_infarction}). 
Note the suggestive inversion of T-wave, a symptom related to myocardial ischemia \cite[]{lin2013electrocardiographic} .
Lastly, for \gls{sbr}, we can appreciate a regular rhythm with larger R-R intervals compared to those of Fig. \ref{fig: example_Sinus_Rythm} (see Fig. \ref{fig: example_Sinus_Bradycardia}).  

\section{Methodology}\label{sec:methodology}

In this section, we present the strategy employed to develop \gls{ann}-based interpretable predictions for detecting various \gls{ecg} irregularities. Our approach integrates a binary classifier, specifically the \gls{1dcnn}, with its corresponding \gls{xai} method, the \gls{1ddnn}, to extract significant predictors. We then utilize a \gls{lr} model to construct interpretable predictions
Such an easy-to-interpret model is supported by a feature selection driven to identify, through the \gls{1ddnn}, the most promising patterns that the \gls{1dcnn} learns to distinguish the \gls{ecg} irregularities of interest.

Briefly, we can summarize our methodology in four main stages:
\begin{enumerate}
\item Firstly, we trained a \gls{1dcnn} with orthogonal convolutional kernels \cite[]{li2019orthogonal, wang2020orthogonal} to predict \gls{ecg} irregularities. 
It is important to note that, due to variations in kernel size across convolutional layers, we refer to this approach as semi-orthogonality regularization, employing semi-orthogonal convolutional kernels. Following the training phase, the \gls{1dcnn} demonstrated its capability to effectively distinguish abnormal \gls{ecg} signals from those in the control group. 
Specifically, each classification task required the fitting of a dedicated \gls{1dcnn} model (see Section \ref{sec:model_fit}).
\item 
Secondly, in order to visualize the patterns captured by the \gls{1dcnn} we used a \gls{1ddnn} (see Section \ref{sec: 1dcnn_inversion}). 
This stage aimed at creating the \gls{sm}s which will reveal the most relevant patterns captured by the \gls{1dcnn} (see Section \ref{sec: 1-D DNN-based saliency maps}).
Note that in order to ensure a proper inversion of the \gls{1dcnn}s through the \gls{1ddnn}, we imposed a semi-orthogonality constraint on the convolutional layers of \gls{1dcnn}.
This step was crucial to finalize the creation of accurate and faithful \gls{sm}s (see Appendix \ref{apx: inversion_1d_convolutional_layer}). 

\item Thirdly, we extracted the \gls{ecg} features to feed the \gls{lr}-based interpretable models. 
This stage was accomplished by analyzing the \emph{most relevant patterns} indicated by the \gls{sm}s, i.e. those patterns supporting at most the \gls{1dcnn} predictions.
However, \gls{sm}s provide specific information per each instance and it may vary by instance.
We then opted for clustering the relevant patterns using the \gls{ks} algorithm \cite{paparrizos2015k}. 
The feature extraction then consisted of determining the \emph{presence} of the K-centroids along the time domain of each instance.
We recall that the K-centroid refers to the refers to the main shape or prototype of a cluster formed during the clustering process.
We anticipate that with the term \emph{presence} we intend a measure of similarity between a K-centroid and a segment of an \gls{ecg} instance. 
The presence of a K-centroid on a segment increases with the closeness of that centroid's match to the segment.
A detailed discussion is reported in Section \ref{sec: feature extraction}.

\item Finally, we implemented a \gls{lr} to make interpretable predictions. 
The \gls{lr} model was fed with features extracted via \gls{sm} analysis. 
We compared the performance of the \gls{lr} and \gls{1dcnn} models to assess if this methodology offers predictions as accurate as \gls{ann} methods while ensuring interpretability.
The intelligibility of the set of extracted features gives the desired interpretability feature to the way of resolving the classification task.
We shall discuss these aspects in Section \ref{sec: interpretable predictions}.
\end{enumerate}
\subsection{Time-series instances}

The instances we used to feed the statistical learning models are derived from the data introduced in Section \ref{sec:data}.
To make the learning phase as robust as possible, we adapted some additional pre-processing steps to the \gls{ecg} data.
Then, we rescaled the \gls{ecg}s signals by a factor $10^{-3}$, made them zero-mean and apodized the signals with Tuckey window in order to make them irrelevant at the boundaries.
Note that the last pre-processing step helps contrast the creation of artifacts when the \gls{1dcnn} processes an instance in proximity to its edges.
We specify that \gls{ecg} data can be assimilated to 1-D time-series.
For each patient, we possess one \gls{ecg} independent of the others.
For convenience, we shall utilize the term \gls{tsi} to refer to the \gls{ecg} record of one generic patient.

\subsection{Designing and Training the 1-D CNN}\label{sec:model_fit}

To solve the binary classification tasks we opted for 
a \gls{1dcnn}.
According to the classification task to solve, the model's architecture may differ in some hyper-parameters.
In general, all models were developed using the following scheme:
\begin{enumerate}
    \item The input \gls{tsi}s were firstly propagated by a convolutional layer.
    A \gls{relu} was then applied after the convolutions. 
    The choice of \gls{relu} is necessary to allow us to invert the model through the original implementation of a \gls{1ddnn}, as introduced by \cite{Zeiler2014}.
    The \emph{activated feature maps} (i.e., the output of the \gls{relu} activation layers) were then down-sampled through a max-pooling layer (pooling size equal to 2 pixels).
    By similarity, we shall refer to the \emph{pooled feature maps} as the output of the pooling layer.
    
    \item The layers scheme (i.e., the sequence of convolution, activation function, and max-pooling layers) was then repeated for an integer number of times.
    We shall denote this model's feature with the term \emph{deepness}. 
    
    \item Finally, the resulting feature maps were flattened through a \emph{flatten layer} and propagated through a \emph{dense layer} with one output unit and sigmoid activation function.
\end{enumerate}

The models optimize the binary categorical entropy loss function; the maximum upper bound of training epochs was set to 10000.
The optimization algorithm we opted for is the ADAM \cite[]{kingma2014adam}; the learning rate was initially set to $10^{-3}$ and was rescheduled after each epoch with a constant decay rate of $10^{-7}$.

As introduced, the 1-D convolutional layers are constrained to be semi-orthogonal.
That is, denoting with $w_{irk}$ the $r$-th pixel of the convolutional kernel mapping the $i$-th input feature in the $k$-th output convolutional feature, the semi-orthogonality constraints reads
\begin{equation}\label{eq: definition_semi_orthogonality_constraint_1dcnn}
    \sum_{r = -R}^{R}\sum_{k = 0}^{K} w_{irk}w_{krl} - \delta_{il} = 0;
\end{equation}
with $\delta_{il}$ the Kronecher's delta.
While deriving the condition of semi-orthogonality, we assumed the 1-D convolutional kernel to have an amplitude of $2R+1$ (where $R$ is a generic integer); we denoted with $K$ the total number of output features.

Note that the semi-orthogonality constraint was implemented on each 1-D convolutional layer, following the algorithm proposed by \cite{povey2018semi}. 
Embedding this constraint within a \gls{1dcnn} requires solving additional optimization problems during the learning phase. 
Although this could introduce some complexity and potentially slow down the training process, the employed algorithm guarantees quadratic convergence, ensuring both fast and precise fulfillment of the desired constraint.

\begin{figure}
    \centering
    \includegraphics[height=9cm]{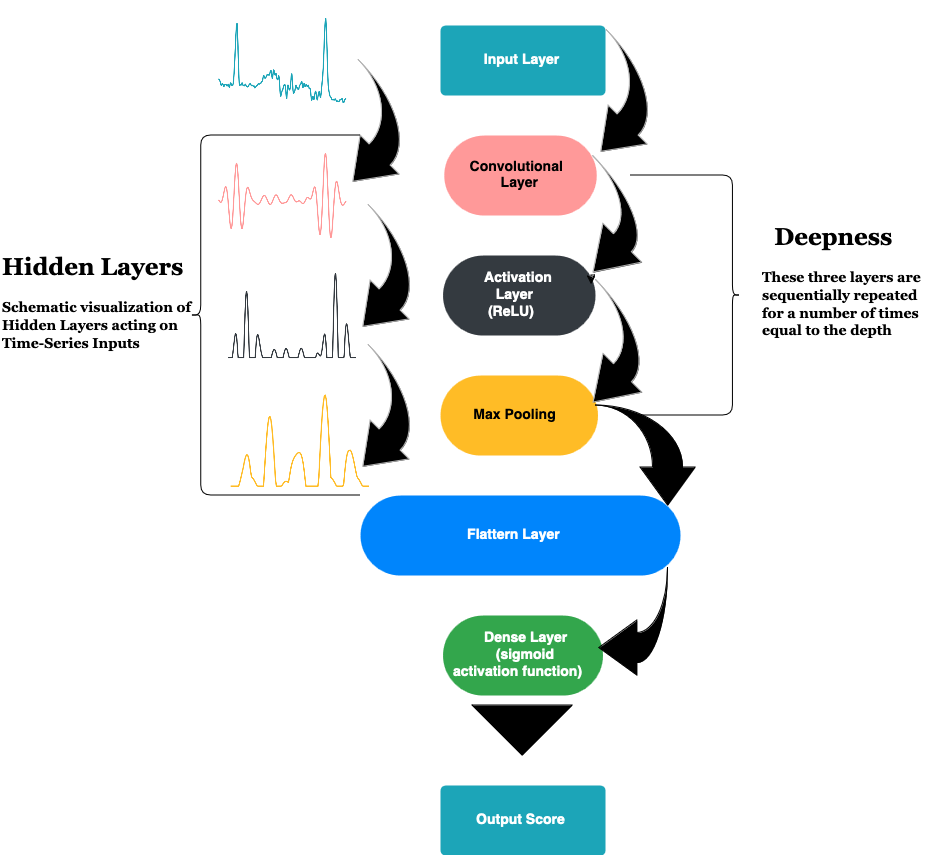}
    \caption{Schematic Representation of the \gls{1dcnn} employed for Binary Classifications of \gls{ecg} data}
    \label{fig: 1dcnn_model}
\end{figure}

Regularization techniques were also implemented to contrast the model's over-fitting.
In particular, we utilized the \emph{early stopping method }.
The training phase was stopped whenever the model's error on the validation set didn't decrease for 5 consecutive iterations.
Unlike other standard techniques such as Dropout \cite[]{srivastava2014dropout} or Gaussian Dropout \cite[]{kingma2015variational, wang2013fast}, the \emph{early stopping method } does not alter the output of the convolutional layers, thereby preserving the imposed semi-orthogonality constraint more effectively.
A schematic figure of the model is presented in Figure \ref{fig: 1dcnn_model}.

\begin{figure}
    \centering
    \includegraphics[width=.5\linewidth]{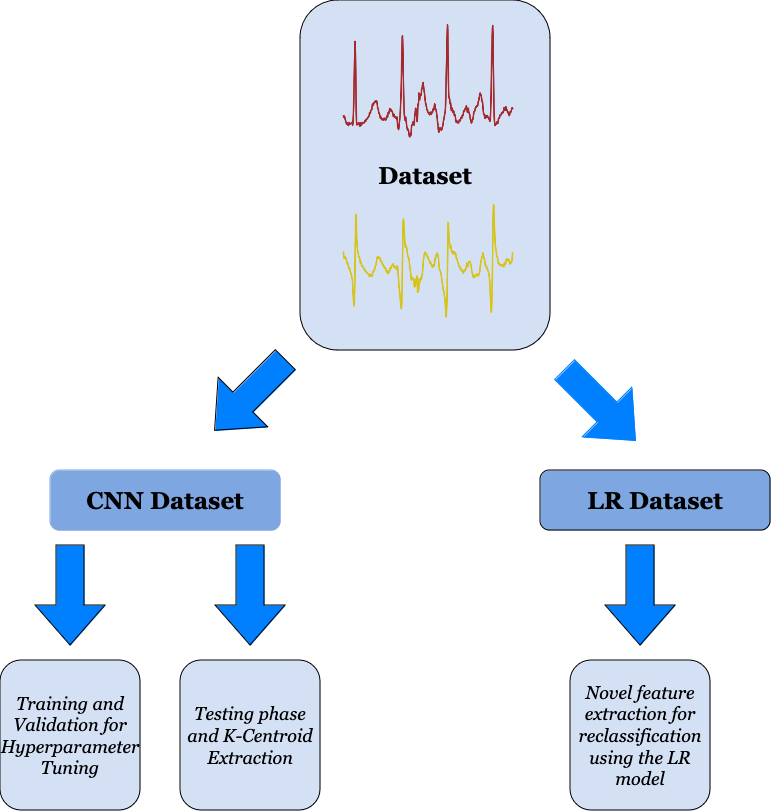}
    \caption{Diagram illustrating the data splitting procedure. An initial split creates the CNN-Dataset and LR-Dataset. The CNN-Dataset is further divided into training, validation, and testing subsets, which are used to train, tune, and validate the \gls{1dcnn}, and subsequently test the model. The testing subset provides the K-centroids for feature extraction. The LR-Dataset is then used to perform a reclassification task with \gls{lr}, where the K-centroid-based features extracted from the LR-Dataset are utilized.}
    \label{fig: datasplit_scheme}
\end{figure}

To make the model's selection and testing, data were split into different sub-datasets.
More precisely, an initial data splitting was accomplished to determine two equal different datasets each one designed to feed either the \gls{1dcnn} or the \gls{lr}.
For clarity, we shall refer to these splits as the \emph{CNN-Dataset} and the \emph{LR-Dataset}, respectively.
The CNN-Dataset was used to train, validate and test the \gls{1dcnn} with the scope of providing the K-centroids to construct an algorithmic method to extract interpretable \gls{ecg} features.
These interpretable features were subsequently extracted from the LR-Dataset and utilized to perform once again a classification task using the \gls{lr} model. 
This step aimed to evaluate whether the data-driven feature extraction could effectively solve the classification problem optimally.
A summary of this process is presented in Figure \ref{fig: datasplit_scheme}.

The search for the best hyperparameters of \gls{1dcnn}s was accomplished through a 5-fold cross-validation. 
Initially, 20\% of the available data was set aside as a fixed \emph{test set}. 
The remaining 80\% was then utilized to train and validate the model. 
We stress that the \emph{test set} remained untouched throughout the training process, ensuring its independence for final evaluation.
Within the 5-fold cross-validation scheme, the \emph{training} and \emph{validation sets} were dynamically selected, allowing each fold to serve as both a \emph{training} and \emph{validation} dataset in turn. 
The model's error was evaluated using the \emph{validation set} from each fold, guiding the early stopping method to prevent overfitting. 

A Grid-search approach was utilized to choose the optimal hyperparameters to explore. 
\gls{1dcnn}s models were trained over different combinations of hyperparameters such as \emph{number of filters}, \emph{convolutional kernel size}, and \emph{deepness}.
The best configuration was chosen after considering three different metrics, i.e., the \gls{auroc}, \gls{auprc}, and the \gls{mcc}; the configuration with the highest product of these metrics was deemed the best. 
In what follows, we shall refer to the product of these metrics as the \gls{3mcs}. 
More specifically, given some generic pairs of predictions $\tilde{y}$ and ground labels $y$, we define it as
$$ \operatorname{3MCS}(\tilde{y}, y)  := \operatorname{AUROC}(\tilde{y}, y)\, \operatorname{AURPC}(\tilde{y}, y) \frac{1+\operatorname{MCC}(\tilde{y}, y)}{2}.$$
Note that \gls{auroc} and \gls{auprc} take values between 0 and 1, while \gls{mcc} ranges in [-1, 1].
Also, \gls{mcc} requires labeled prediction, while the \gls{1dcnn} produces a score between 0 and 1.
To overcome this issue, we calibrated the model through the \emph{isotonic regression} \cite{de1977correctness}.
As a result, the prediction labeling was then based on the adjusted predicted probabilities. 
All these measures share a common characteristic: the closer their value is to one, the more accurate the model's performance.
\gls{auroc} gives general performance measures and it is usually adopted as a standard metric to asses models' performances in many classification contexts.
\gls{auroc} can be misleading for problems with low sensitivity and specificity, potentially leading to the drawing of optimistic conclusions. 
In such cases, \gls{mcc} represents a more valid option, as high \gls{mcc} values are always associated with high \gls{auroc}, and never vice-versa \cite[]{chicco2023matthews}.
\gls{auprc} is preferred for imbalanced datasets (i.e., the majority of the dataset belongs to one specific class), as it more accurately focuses on positive class performance compared to \gls{auroc}. 

As emerged from Section \ref{sec: Datasets} (see table \ref{tab: Summary_Statistic}), all 3 classes of irregular \gls{ecg} are in a vast minority with respect to the rest of \gls{ecg}. 
As a result, the imbalance of datasets might dramatically affect the resolution of binary classification tasks leading either to poor or too optimistic results.
For this reason, while training and validating any statistical learning model, we opted for contrasting data imbalance through the \gls{smoteenn}; a technique presented in \cite{batista2004study} to make the learning dataset balanced through the sequential combination of \gls{smote} \cite[]{chawla2002smote} and \gls{enn} \cite[]{wilson1972asymptotic}.
Incidentally, \gls{smoteenn} has been already successfully used in \gls{ecg} classification problems, e.g. see \cite{manju2019classification},\cite{dixit2021early}, and \cite{mandala2024improved}. 
Specifically regarding the \gls{smoteenn} methodology, \gls{smote} is a \gls{knn}-based over-sampling technique, creating new instances along the line connecting two random first-neighbours from the minority group. 
Such a technique might create artefacts, for this motivation the data balance of \gls{smote} is refined by means of \gls{enn}; a \gls{knn}-based technique to remove instances which does not match the majority class label among its neighbors.

Hence, the optimal configuration for the three binary tasks is as follows: 32 convolutional filters, kernel size of 5 pixels, and a depth of 2 for the \gls{af} task; 32 convolutional filters, kernel size of 5 pixels, and a depth of 3 for the \gls{mi} task; and 32 convolutional filters, kernel size of 9 pixels, and a depth of 4 for the \gls{sbr} task. A comprehensive comparison of model performance across these different configurations is provided in Appendix \ref{apx: hyperparamenters}, offering a detailed view of the tuning process.

\subsection{Inversion of 1-D CNN}\label{sec: 1dcnn_inversion}

In order to understand the mechanism of the forward propagation underlying the layers of the \gls{1dcnn}, we opted for implementing a \gls{1ddnn}.
In this work, we adopted the strategy proposed by \cite{zeiler2010deconvolutional} and \cite{Zeiler2014}.
Thus, we constructed a \gls{1ddnn} to attempt reconstructing the inverted mapping from the deepest layers of the \gls{1dcnn} back to the input space.
More specifically, the reconstruction was meant from the output of the deepest max pooling layer of the \gls{1dcnn}.

\begin{figure}
    \centering
    \includegraphics[width= \textwidth]{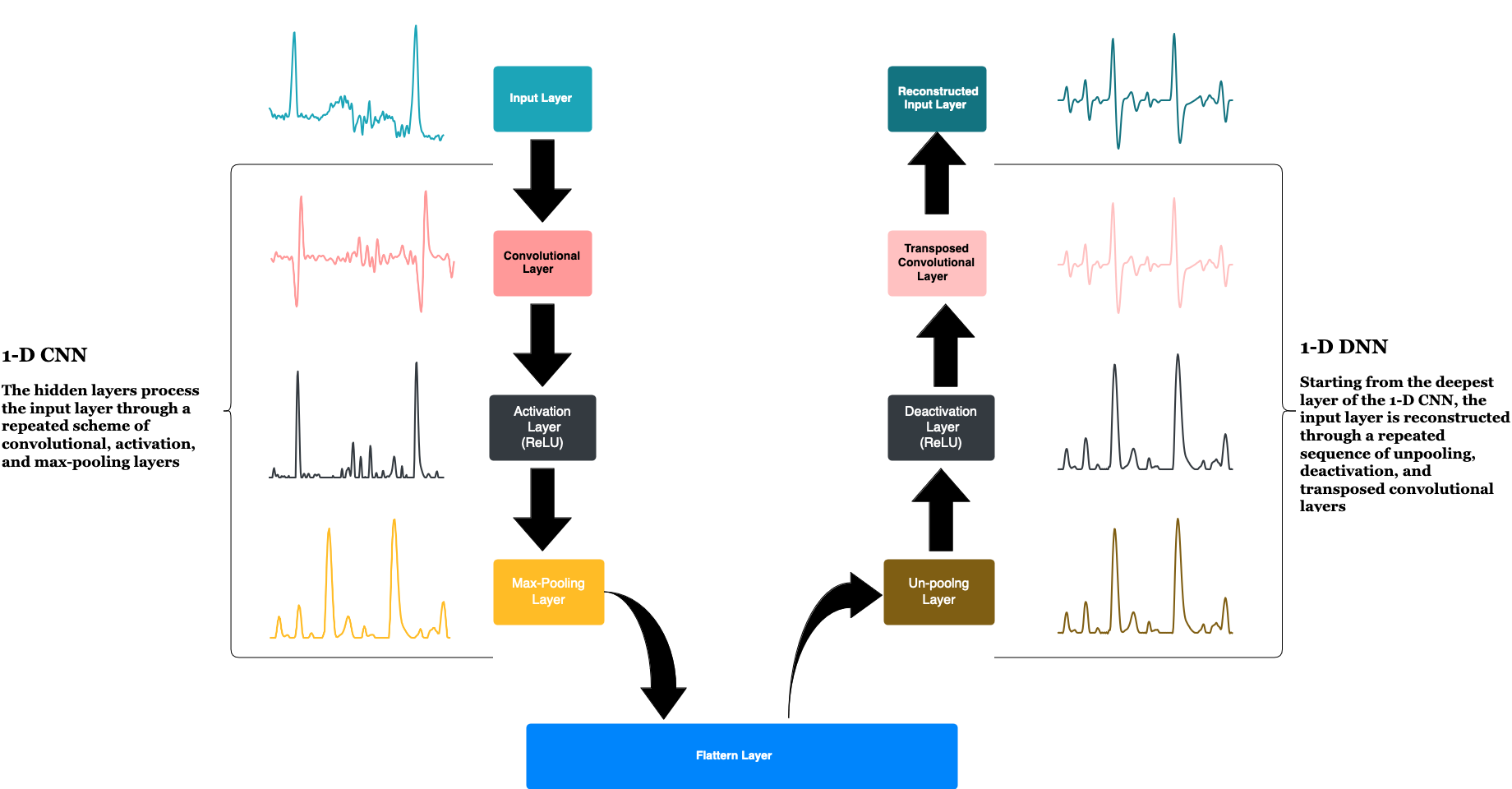}
    \caption{Scheme of a \gls{1ddnn} applied to the hidden layers of a \gls{1dcnn}}
    \label{fig: scheme_1ddnn}
\end{figure}

A \gls{1ddnn} is a sequence of non-trainable algorithmic layers that are arranged with the scope of sequentially inverting the hidden layers of a \gls{1dcnn}.
Those non-trainable layers are constructed from the parameters trained in the hidden layers of the \gls{1dcnn}. 
A scheme of \gls{1ddnn} is reported in Figure \ref{fig: scheme_1ddnn}.
%

As already mentioned, we used the \gls{1dcnn} in order to invert a sequence of convolutional, activation functions, and max pooling layers.
A convolutional layer is naturally inverted by a \emph{transposed convolutional layer}, which is constructed by transposing the convolutional weights of the original layer.
Activation functions are inverted through their natural inverse function when possible. 
Despite the ReLU is a non invertible function,
as stated by \cite{Zeiler2014}, the ReLU itself can be utilized as an inverse activation function.
This strategy should help the \gls{1ddnn} keep locations occluded by the activation functions and reveal information irrelevant to the final prediction.
The pooling layers represent another source of non-linear operators that do not admit an inverse transformation.
However, an inversion could be attempted through the method of \emph{Switch Variables}; see \cite[]{Zeiler2014}.
The method of the Switch Variables consists of preserving the original configuration of maxima before the pooling operation takes place; during the reconstruction, the maxima are reassigned to their original locations.
This strategy gives better reconstructions compared to the unpooling via bilinear interpolations.
However, the formulation of the method does not give any directions on the values to assign to non-maximal locations.
In our applications with a 2-sample pooling, we utilized linear interpolation while ensuring the reconstructed records were positive and less than the pooled maximum. 

In the case of convolutional layers, it is important to note that while the input of the convolutional layers can be approximated using transpose convolutional networks, as described by \cite{zeiler2010deconvolutional} and \cite{Zeiler2014}, an exact match between the original input and the reconstructed one is not always guaranteed. 
However, this reconstruction becomes exact when the semi-orthogonality constraint is applied to the convolutional layers. 
The mathematical proof supporting this is provided in the Supplementary materials (For 1-D convolutional layers see Appendix \ref{apx: inversion_1d_convolutional_layer}; a generalization for the 2-D convolutional Layer and the Dense Layer are reported in Appendix \ref{apx: inversion_2d_convolutional_layer} and \ref{apx: inversion_dense_layer}, respectively).

\subsection{1-D DNN-based saliency maps}\label{sec: 1-D DNN-based saliency maps}

\gls{sm}s are powerful mathematical tools to quantify the degree of importance that a \gls{ann} model gives to the patterns located along the whole spatial domain of a generic \gls{tsi}.
The scope of \gls{sm}s is to measure the extent to which patterns within a particular area of the spatial domain contribute to the model's output.
When dealing with classification tasks, the saliency maps are usually designed to point out the patterns mostly captured by the model to make the predictions accurate at best.

In this section, we propose a method for constructing a \gls{1ddnn}-based \gls{sm}s.
The main mechanism of this method is based on understanding the correspondence between the input data and the deepest max-pooling layer in a \gls{1dcnn}.
More specifically, we aim to explain the connection between the input data and the \emph{deepest latent representations} provided by the \gls{1dcnn} immediately before forwarding them through the final 1-output dense layer (i.e., the prediction layer). 

Let us denote a generic \gls{tsi} with $X$. 
Initially, we propagate $X$ through a fitted \gls{1dcnn}, extracting the \emph{deepest latent representation}, i.e., the output from the deepest max-pooling layer. 
This output is then forwarded through the \gls{1ddnn} (as introduced in Section \ref{sec: 1dcnn_inversion}) to reconstruct $X$; we denote the reconstruction with $X'$. 
Evaluating this reconstruction is crucial for determining the saliency, denoted as $\varphi$. 
As anticipated, the saliency measures the importance of a pattern for the model's predictions. 
In binary classification problems, \gls{sm} should reflect the ability of patterns to be either occluded or propagated as they pass through the activation and max-pooling layers.
In other words, we assume that the reconstruction provided by the \gls{1ddnn} should be able to faithfully reproduce those relevant patterns that the hidden layers would not occlude.
Conversely, patterns that would be occluded and thus not salient for the \gls{1dcnn} cannot be successfully reconstructed. 
In light of this, we propose a \gls{sm} based on the discrepancy between $X$ and $X'$, namely
\begin{equation}\label{eq: saliency_maps_chi_squared_}
    \varphi(Z) = 1-\frac{1}{\Gamma(1/2)}\gamma(1/2, Z^{2});
\end{equation}
where $Z$ denotes the standardization of the point-wise discrepancy $X-X'$, $\Gamma$ the Euler's Gamma function, and $\gamma$ the lower incomplete Gamma function.
We shall refer to the map \eqref{eq: saliency_maps_chi_squared_} as the \gls{cssm}.
Note that for a one \gls{tsi}, $\varphi$ takes the form of a matrix; by convention, the saliency maps are located along the rows, and the number of rows is equal to the number of features.
By assuming that $Z \sim \mathcal{N}(0, 1)$, the \eqref{eq: saliency_maps_chi_squared_} descends directly from the survival function of a Chi-squared distribution with one degree of freedom.
Accordingly, the closer $Z$ is to 0, the larger is the chance that the reconstruction $X'$ is faithful to $X.$
By reflection, the closer $\varphi$ is to 1, the higher the chance that the reconstruction $X'$ retains a pattern from $X$ that is essential for the model's prediction.
Mathematical details regarding the construction of the \gls{cssm} are provided in Appendix \ref{apx: Chi-squared Saliency Map}.

The assessment of the effectiveness of the saliency maps was investigated through two validation steps.
We verified that the reconstructed \gls{tsi}s are still classified accurately when propagated through the model. 
This step is essential to ensure that the reconstructed \gls{tsi}s still possesses a piece of predictive information coming for the original \gls{tsi}s.
Next, we evaluated the quality of the saliency maps using the methods \gls{roar} \cite[]{mundhenk2019efficient}.
Such a method was initially proposed for explaining \gls{2dcnn}, but a readaptation of such a method for \gls{1dcnn} has been already employed; see \cite{lancia2022physics}.  
In this step, we iteratively occluded increasing quantiles of the most salient spatial locations in the input data and then measured the model's accuracy.
The expectation is that the accuracy will decrease as the proportion of occluded pixels grows.

\subsection{Saliency maps-based feature extraction}\label{sec: feature extraction}

In order to construct a set of new explainable features, we utilized the information provided by \gls{cssm} about the pattern recognition activity of the \gls{1dcnn}.
More specifically, we inspected the \gls{sm}s to identify and extract the segments with the \emph{most salient patterns} of each \gls{tsi}.
By "most salient patterns," we refer to the input domain with the largest averaged saliency.
Consequently, the extraction of saliency map-based features depends on the size of the chunk investigated.
Selecting the most salient patterns allows us to focus only on time-series chunks that have a major affinity with the model's predictions.

Since the single \gls{sm} provides specific information for one \gls{tsi}, the most salient pattern is exclusive to that \gls{tsi}. 
To identify more general salient patterns, we employed a clustering method to find the most representative candidates that describe the salient patterns across different \gls{tsi}s.
Thus, we opted for the \gls{ks} method; an unsupervised clustering algorithm designed to group time series data based on shape similarity \cite[]{paparrizos2015k}.
We adopted the centroids of the \gls{ks} as the set of candidates that generalize the most salient patterns among all \gls{tsi}s.

We based our feature extraction on the \emph{presence} of the \gls{ks} centroids within each \gls{tsi}.
With the term presence, we refer to the ability of \gls{ks} centroids to match with any segment of one \gls{tsi}.
The presence of a \gls{ks} centroid was evaluated by sliding it along the \gls{tsi} and taking the L1 norm at each position. 
We just took the L1 norm with the minimal value. 
More specifically, given $X$ a generic \gls{tsi} and $\mathfrak{K}$ a \gls{ks} centroid, the presence of $\mathfrak{K}$ in $X$ is evaluated as 
$$\min_{t} ||X_{t\to t+L}-\mathfrak{K}||_{1};$$
with $L$ the amplitude of the time domain of $\mathfrak{K}$ and $X_{t\to t+L}$ the segment of $X$ extracted from $t$ to $t + L.$
This procedure is inspired by the learning process of the Shapelets model \cite[]{ye2009time}, which optimizes the shape of local patterns (i.e., shapelets) within time series data that are most indicative of solving a classification task.
In order to increase the separability of these features, we utilized a \gls{kpca} with a radial basis function kernel.
The number of \gls{kpca} components was equal to the number of the extracted features.

\subsection{Interpretable predictions}\label{sec: interpretable predictions}

The previous sections mainly focused on making the \gls{1dcnn} learning phase explainable. 
As introduced, the \gls{1dcnn}-based \gls{sm}s aim to reveal the patterns in the \gls{tsi}s that are most effective for the classification task. 
In Section \ref{sec: feature extraction}, we introduced a feature extraction method based on the insight revealed by the \gls{1dcnn}-based \gls{sm}s. 
This method provides a specific meaning to the extracted features based on the presence of some candidate temporal patterns within the \gls{tsi}s.
Therefore, we leveraged this characteristic to make interpretable predictions for the classification tasks analyzed by the \gls{1dcnn}.
To achieve this, we will use the extracted features to feed a \gls{lr}.

To fit the \gls{lr}, we utilized data that was neither involved in the training phase nor in the testing phase of the \gls{1dcnn}.
We recall that before training the \gls{1dcnn}, the data was randomly split in half, with one half used for training, validating, and testing the \gls{1dcnn}, and the other half reserved for constructing the interpretable \gls{lr}-based prediction model.
Both training and validation phases of \gls{lr} had the scope of revealing whether its prediction power is as accurate as that of the \gls{1dcnn}. 
For this, we compared the relative errors of the point estimates of \gls{auroc}, \gls{auprc}, and \gls{mcc} between the two models. 
Additionally, a feature inspection of the \gls{lr} model helped assess whether the model could effectively base its predictions on meaningful features. 
As a feature inspection method, we opted for the \emph{Feature Importance}, assigning an importance degree to each feature equal to the relative model's error. 
Note that a direct application of the Feature Importance will outline the importance of the \gls{kpca} components. 
To show the underlying relation between the most important feature and the presence of the \gls{ks} centroids, we utilized Spearman's correction test.
A posteriori visualization of the \gls{ks} centroids mostly correlated with the most important \gls{kpca} feature helped determine whether the predictions were actually based on known characteristics of \gls{af}, \gls{mi}, or \gls{sbr}.

\section{Results}\label{sec:results}

\subsection{Assesment of 1-D CNN predictions}
Optimal configurations for the \gls{1dcnn} were selected using grid-search validation, where the best configuration was chosen based on the highest performance metrics identified during this phase.
For the \gls{af} dataset, the optimal configuration consisted of 32 filters, kernel size of 5, and deepness equal to 4; for the \gls{mi} dataset, 32 filters, kernel size of amplitude 5, and deepness of 3; for the \gls{sbr}, 32 filters, kernel size of 9, and a deepness of 4.
With those configurations, we observed that the highest \gls{3mcs} was equal to $0.83 \pm 0.02$ for the \gls{af}, $0.99 \pm 0.01$ for the \gls{mi}, and $0.93 \pm 0.01$ for the \gls{sbr}.
Tables \ref{tab: Model's Selection AF}, \ref{tab: Model's Selection MI}, and \ref{tab: Model's Selection SB} provide a complete overview of the metrics evaluated during the validation phase; see Appendix \ref{apx: model's selection}.

The testing phase revealed results similar to those of the validation phase.
For the mentioned optimal configurations; for \gls{af}, we observed
an \gls{3mcs} equal to 
$0.85 ± 0.01$;  for \gls{mi} , $0.99 ± 0.01$; for \gls{sbr} $0.83 ± 0.01.$
Specifically; for the \gls{af} dataset we observed \gls{auroc} equal to $0.95 \pm 0.01$, \gls{auprc} $0.98 \pm 0.01$, and $\gls{mcc}$ of $0.77\pm0.01.$; 
for the \gls{mi} we found \gls{auroc} equal to $0.99\pm0.01$, \gls{auprc} equal to $0.99 \pm 0.01$, and \gls{mcc} equal to $0.99\pm0.01;$ 
for the \gls{sbr} we then observed \gls{auroc} $0.95\pm0.01$, \gls{auprc} $0.94\pm0.01$, and \gls{mcc} equal to $0.85 \pm 0.01.$
A summary of these results is shown in Table \ref{tab: test_set_performace_metrics}.
\begin{table}[]
    \centering
    \begin{tabular}{lcccc}
    \toprule
         & AUROC &  AUPRC & MCC & 3MCS \\
    \midrule
    AF Data & $0.95 \pm 0.01$ & $0.98 \pm 0.01$ & $0.77 \pm 0.01$& $0.82 \pm 0.01$\\
    MI Data & $0.99 \pm 0.01$ & $0.99 \pm 0.01$ & $0.99 \pm 0.01$& $0.99 \pm 0.01$\\
    SB Data & $0.95 \pm 0.01$ & $0.94 \pm 0.01$ & $0.85 \pm 0.01$& $0.83 \pm 0.01$\\
    \bottomrule
    \end{tabular}
    \caption{Performance metrics for the test set }
    \label{tab: test_set_performace_metrics}
\end{table}

\subsection{Comparision with LSTM-based predictions}\label{sec: LSTM-based predictions}
For completeness, the \gls{1dcnn} model was also compared with other two \gls{ann} models, i.e., the \gls{lstm} and the \gls{1dcnnlstm} model.
With the latter, we mean a \gls{1dcnn}, as presented in Section \ref{sec:model_fit}, where the Flatten layer is substituted with a \gls{lstm} layer.
Also, when finding the best configuration of the \gls{1dcnnlstm}, we adopted the best configuration obtained for the \gls{1dcnn} and we consider to optimize the number of \gls{lstm} units.
Likewise, the search for the optimal configuration of the \gls{lstm} consisted of optimizing the number of \gls{lstm} units.
We utilized the same validation and testing approach utilized for the case of \gls{1dcnn}.
For the \gls{af}, the optimal number of \gls{lstm} units was equal to 32 and 16 for the \gls{lstm} and \gls{1dcnnlstm}, respectively.
For the \gls{mi}, the optimal number of \gls{lstm} units was equal to 64 in both models.
For the \gls{sbr}, the optimal number of \gls{lstm} units was equal to 8 and 2 for the \gls{lstm} and \gls{1dcnnlstm}, respectively.
A full presentation of the evaluations accomplished during the validation phase of both models is reported in Appendix \ref{apx: model's selection}.

When those configurations were tested, we observed pretty different results. 
Starting from the \gls{lstm}, we obtained, for \gls{af}, a \gls{3mcs} equal to $0.41 \pm 0.18$; for \gls{mi}, a \gls{3mcs} equal $0.63 \pm 0.11$; for \gls{sbr}, a \gls{3mcs} equal $0.59 \pm 0.15.$
For the \gls{1dcnnlstm}, we obtained, for \gls{af}, a \gls{3mcs} of $0.72 \pm 0.04$, for \gls{mi}, a \gls{3mcs} of $0.95 \pm 0.02$; for \gls{sbr}, a \gls{3mcs} of $0.98 \pm 0.01.$
Thus, the \gls{lstm} model showed a much lower skill in solving the classification of irregular \gls{ecg}s compared to the \gls{1dcnn}.
Conversely, the \gls{1dcnnlstm} offered a more sharpened ability in identifying \gls{ecg}s; for \gls{sbr} predictions resulted to be more accurate compared to the \gls{1dcnn}.
In the other cases, however, the predictive power of \gls{1dcnnlstm} appeared to be similar or slightly lower to the \gls{1dcnn}, even though with larger error bars.
Hence, the \gls{1dcnn} showed better results for the majority of the classification tasks, so we selected it as the classification model.

\subsection{Assessment of prediction based on 1-D DNN-derived reconstructions}
\begin{table}[]
    \centering
    \begin{tabular}{lcccc}
    \toprule
         & AUROC &  AUPRC & MCC & 3MCS \\
    \midrule
    AF Data & $0.94 \pm 0.01$ & $0.95 \pm 0.01$ & $0.77 \pm 0.01$& $0.79 \pm 0.01$\\
    MI Data & $0.99 \pm 0.01$ & $0.99 \pm 0.01$ & $0.95 \pm 0.01$& $0.95 \pm 0.01$\\
    SB Data & $0.96 \pm 0.01$ & $0.94 \pm 0.01$ & $0.79 \pm 0.01$& $0.81 \pm 0.01$\\
    \bottomrule
    \end{tabular}
    \caption{Performance metrics for the \gls{1ddnn}-based reconstructions}
    \label{tab: reconstructions_performace_metrics}
\end{table}
When considering the \gls{tsi}s reconstructed through the \gls{1ddnn}, we found that these reconstructions can still effectively solve the classification task. 
The reconstructed test set showed an \gls{3mcs} of $0.79 \pm 0.01$ for the \gls{af}; $0.95 \pm 0.01$ for the \gls{mi}; and $0.81 \pm 0.01$ for the \gls{sbr}; see table \ref{tab: reconstructions_performace_metrics}.
Compared to the three models' performance, we observed an absolute relative error of the \gls{3mcs} equal to $3.5\%$, for \gls{af}; $4.0\%$, for the \gls{mi}; $2.4\%$; for the \gls{sbr}. 

\begin{figure}
    \centering
    \includegraphics[width= \textwidth]{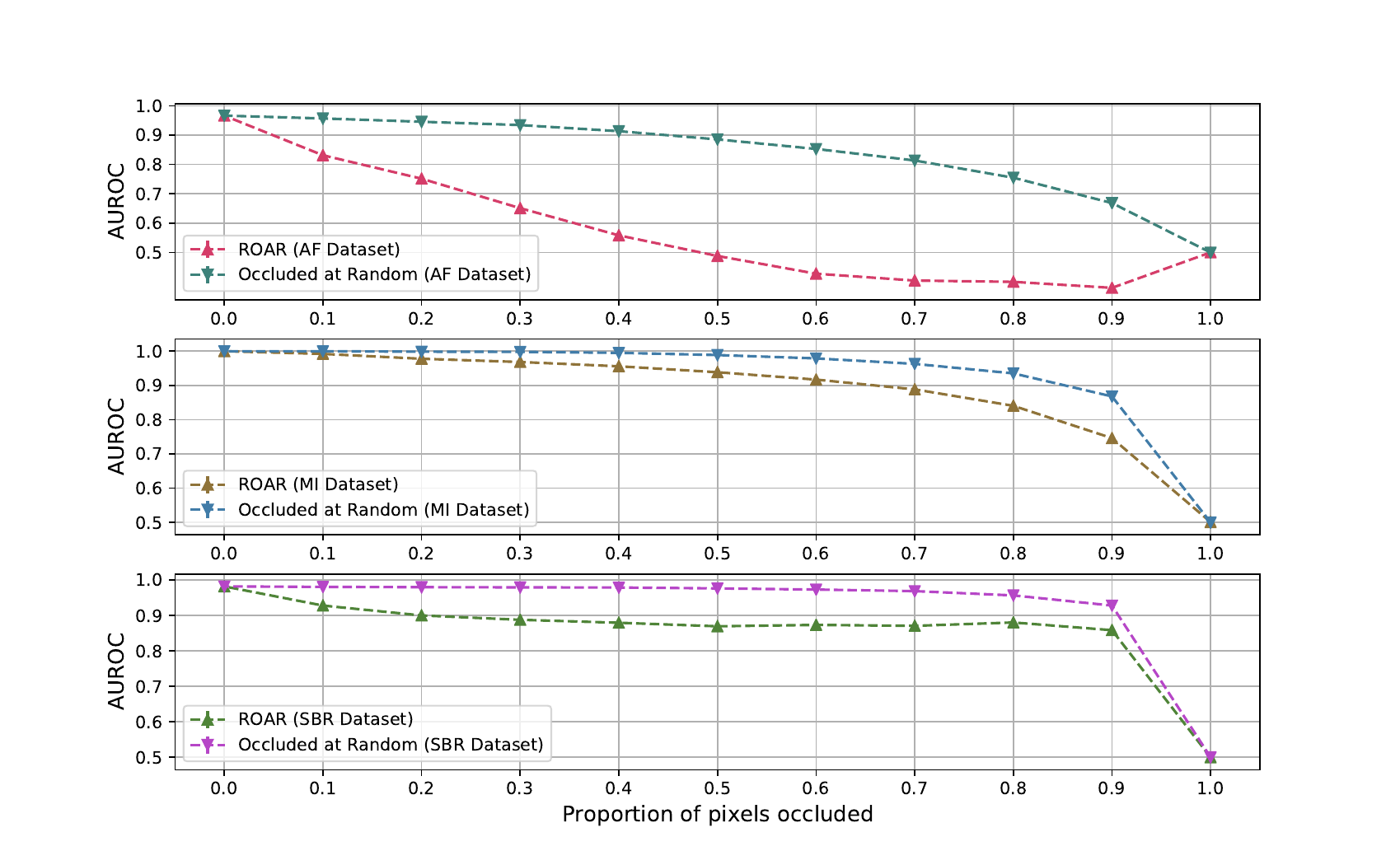}
    \caption{AUC metric as a function of the percentage of pixel occluded. From top to bottom subplots show ROAR for \gls{af}, \gls{mi}, and \gls{sbr}}
    \label{fig: ROAR_ECG}
\end{figure}
The \gls{roar} approach revealed that \gls{sm}s can effectively identify both the most crucial locations and their importance in supporting predictions.
As higher percentage salient pixels, as identified by the \gls{1ddnn}-based \gls{sm}s, are occluded, the predictions (expressed in terms of \gls{auroc}) tend to deteriorate.
Such a deterioration was faster compared to an approach based on random occlusion.
This result was confirmed for all three cases considered; see Figure \ref{fig: ROAR_ECG}.

\subsection{Assessment of interpretable LR-based predictions}
About the \gls{lr} method, we found that a feature extraction based on 32 K-shapes with a time domain amplitude of one second, made the \gls{lr} able to solve the classification tasks for \gls{af}.
Instead, 32 K-shapes with a time domain amplitude of two seconds were necessary for solving the classification task for both \gls{mi} and \gls{sbr}.
Across all cases, the accuracy level of predictions was similar to that of \gls{1dcnn}.
Specifically, for the \gls{af}, the \gls{lr} showed a \gls{3mcs} of $0.74 \pm 0.01 $.
The \gls{mi} case showed an \gls{3mcs} of $0.93 \pm 0.01 $, while the \gls{sbr} a \gls{3mcs} of $0.97 \pm 0.01.$
Compared to the metrics obtained during the testing of the \gls{1dcnn}, we observed, for \gls{af}, relative absolute errors of \gls{3mcs} equal to $2.8\%$; for \gls{mi} $7.1\%$; for \gls{sbr} $12.7\%$.
Note that the result of \gls{sbr} denotes an increase of \gls{3mcs} compared to that achieved by \gls{1dcnn}.
For a complete sensitive analysis regarding the optimal choice of the number of K-shapes adopted during the feature extraction, the analysis of time scales of the K-shapes themselves, and the regularization parameters of \gls{lr} the readers can refer to appendix \ref{apx: sensitivity analysis feature extraction}.

\section{Inspecting Interpretable Predictions}\label{sec: Inspecting_Interpretable_Predictions}
The inspection of the \gls{lr} predictions helped visualize what kind of \gls{ecg} patterns can play a significant role in solving efficiently the various classification tasks.
The Feature Permutation method was utilized with the scope of revealing which extracted features (and which \gls{ks} centroid associated with them) mainly contributed to the final predictions.
We recall that those features were pre-processed through the \gls{kpca} method.
To find the connection underlying the features based on the presence of the \gls{ks} centroids and those pre-processed through the \gls{kpca}, we utilized the Spearman test.
A complete discussion about this step is fully discussed in appendix \ref{apx: correlation_presence_KPCA_features}.

\begin{figure}
    \centering
    \includegraphics[height=0.75\textheight]{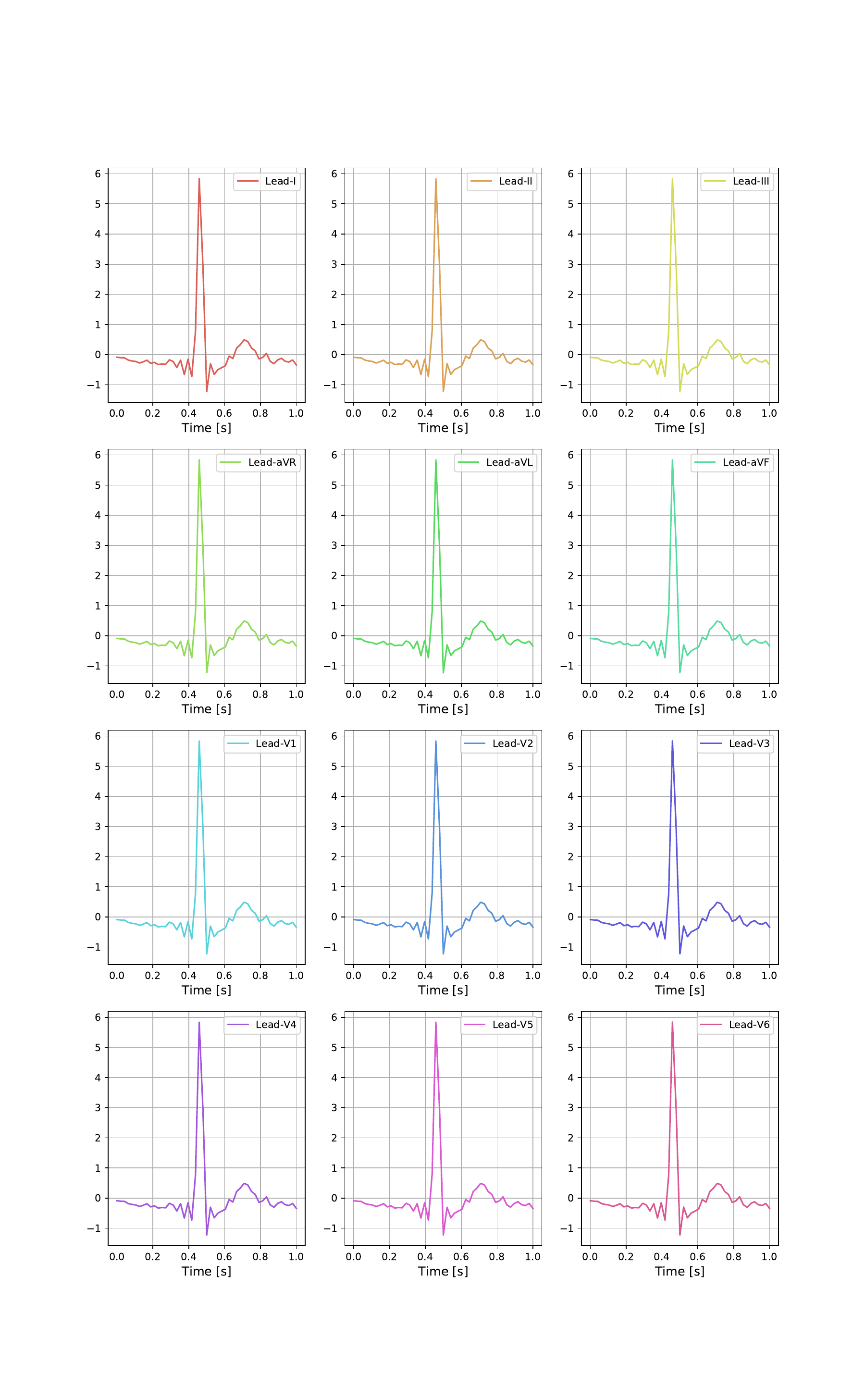}
    \caption{Visualization of the most relevant K-shape for the \gls{af} classification}
    \label{fig: most_relevant_pattern_AF}
\end{figure}
The visualization and inspection of the profiles of the most predictive \gls{ks} centroid, for each of the three classification problems, confirmed that the interpretable predictions are based on true medical characteristics.
For the \gls{af} case, we observed that the most predictive \gls{ks} centroid did not present a regular and well-formed P-wave; it appears to be absent.
All other structures, such as the QRS complex as well as the T wave appeared regular;  see Figure \ref{fig: most_relevant_pattern_AF}.
Such a pattern was present in each of the 12-lead.
We recall that the absence of the P-wave in a standard 12-lead \gls{ecg} represents one main feature to diagnose \gls{af} \cite{hindricks20212020}.

\begin{figure}
    \centering
    \includegraphics[height=0.75\textheight]{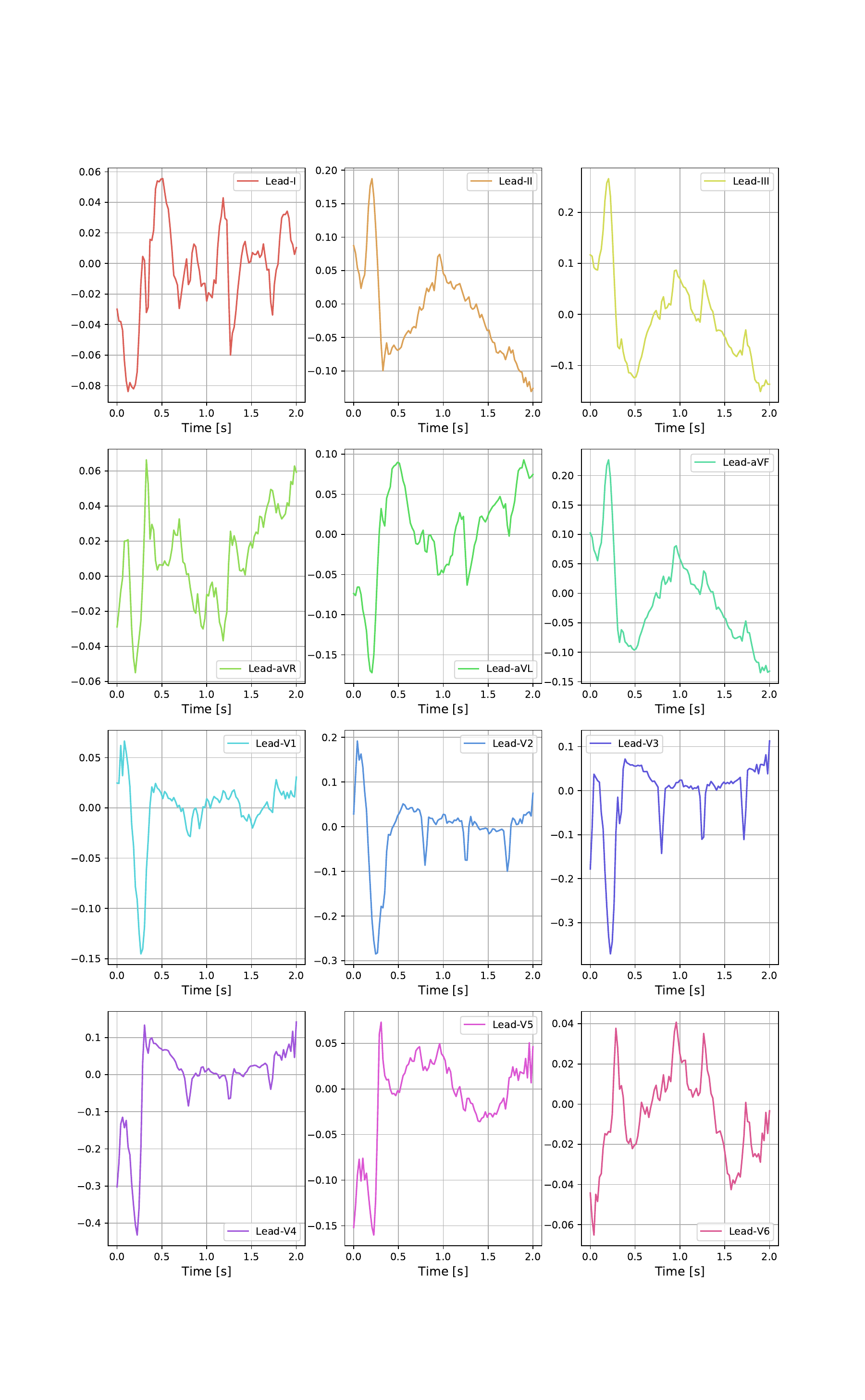}
    \caption{Visualization of the most relevant K-shape for the \gls{mi} classification}
    \label{fig: most_relevant_pattern_MI}
\end{figure}
For the \gls{mi} the interpretation of the most relevant \gls{ks} centroids appeared much more intricate and difficult to lead back to a straight clinical scenario describing \gls{mi}.
In Figure \ref{fig: most_relevant_pattern_MI}, we observed that a possible ST segment depression was depicted for leads II, III, and avF; this is often related to non-ST elevation myocardial infarction \cite{collet2020ten}.
In fact, the S and T points were included in a wave oscillation that appeared to be depicted much lower with respect to the R and Q points; the T wave also seemed to be much wider and more ample.
A similar scenario is also present in lead V6 with an ample, but tall T-wave.
We recall that tall T-wave represents a well-known \gls{ecg} feature associated with the early stages of \gls{mi} \cite[]{dressler1947high}.
Interestingly, in leads V1, V2, V3, V4, and V5 the R-wave appears in opposition to Q-wave.
For the case-specific leads V1-V4, we observed that the T-wave also appread inverted; we recall T-inversion is another characteristic of associtated to \gls{mi} \cite[]{lin2013electrocardiographic}.
Specifically, a group of 3 or four downward peaks is included after the R-wave inversion; we considered the possibility that such a train of inverted T-wave might be an artefact that helps the model identify a T-wave inversion occurring after the QRS complex.

\begin{figure}
    \centering
    \includegraphics[height=0.75\textheight]{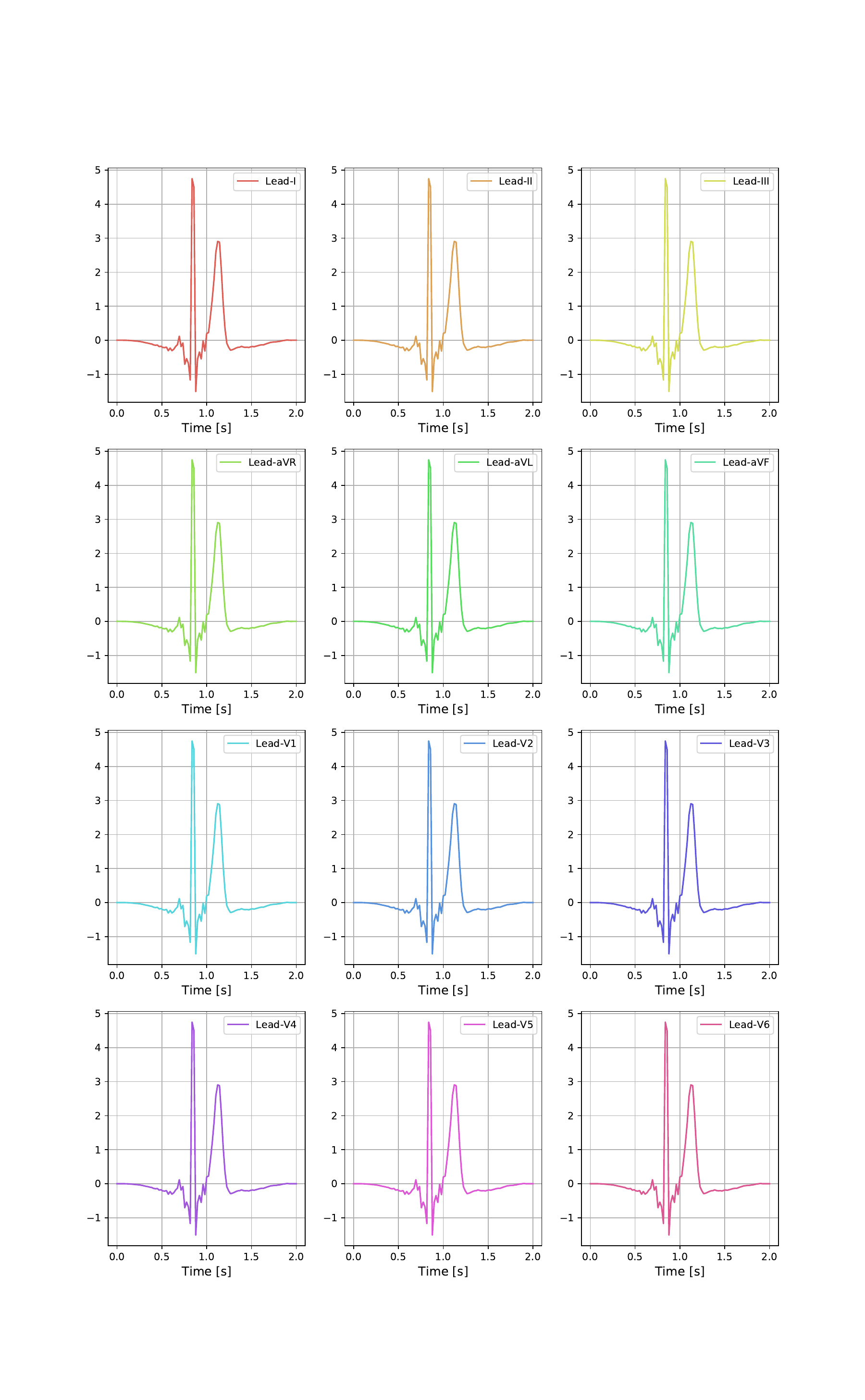}
    \caption{Visualization of the most relevant K-shape for the \gls{sbr} classification}
    \label{fig: most_relevant_pattern_SB}
\end{figure}
For the \gls{sbr} case, we observed a single well-formed and regular QRS complex with no abnormalities with P and T waves; see Figure \ref{fig: most_relevant_pattern_SB}.
As known, \gls{sbr} is characterized by a regular, but slower sine rhythm (a sinus rate lower than 50 bpm) \cite[]{KUSUMOTO2019932}.
Note that the \gls{ks} centroid of Figure \ref{fig: most_relevant_pattern_SB} represents a single beat in a time interval of 2 seconds, where a couple of beats were expected for a normal scenario.

\section{Comparing Interpretable Predictions}\label{sec: Comapring_ Interpretable_Predictions}

In addition to the \gls{lr} model presented in Section \ref{sec: Inspecting_Interpretable_Predictions}, we tried to explore a refinement of it where \gls{ecg}-derived features with available baseline variables, such as sex and age are combined.
This comparison is motivated by our desire to make our study more complete; we want to assess whether integrating a piece of baseline information to the \gls{ecg}-derived features can improve predictive accuracy and enhance the clinical relevance of the \gls{lr} model’s outputs when interpreting cardiac data.

When focusing exclusively on \gls{ecg}-derived features, we can gain insights into specific electrical patterns that may indicate \gls{af}, \gls{mi}, or \gls{sbr}. 
However, while this approach provides valuable information, it may lack the context needed for a comprehensive assessment of a patient's cardiovascular health. 
Incorporating baseline variables, such as age and sex, into our analysis enables a more complete modelling of the risk of the occurrence of irregular patterns of \gls{ecg}.
This dual focus not only aims to improve the accuracy of our predictions but also enhances the trustfulness of the model.

Thus, when comparing the metrics obtained for the full model, we observed that integrating a piece of baseline information results in a slight increase with respect to the \gls{ecg}-derived features model.
For the \gls{af}, we obtained \gls{3mcs} equal to $0.92\pm0.01$, \gls{auroc} $0.98\pm 0.01$, \gls{auprc} $0.98\pm 0.01$, and \gls{mcc} $0.93 \pm 0.01.$
For the \gls{mi}, we obtained \gls{3mcs} equal to $0.94\pm0.01$, \gls{auroc} $0.98\pm 0.01$, \gls{auprc} $0.98\pm 0.01$, and \gls{mcc} $0.93 \pm 0.01.$
For the \gls{sbr}, we obtained \gls{3mcs} equal to $0.98\pm0.01$, \gls{auroc} $0.98\pm 0.01$, \gls{auprc} $0.98\pm 0.01$, and \gls{mcc} $0.98 \pm 0.01.$
Regarding only the \gls{3mcs}, we found an absolute relative error, with respect to the model of Section \ref{sec: Inspecting_Interpretable_Predictions} equal to 8\% for \gls{af}; for \gls{mi} 4\%; for \gls{sbr} 4\%.  

To further substantiate our results, we employed the \gls{lrt} to check whether there was a significant difference between the model using only the \gls{ecg}-derived features and the model incorporating both \gls{ecg}-derived features and baseline covariates. 
This is a widely recognized test used for comparing nested models to determine whether the inclusion of additional parameters leads to a significantly better fit; for example, see Chapter 10 of \cite{wasserman2006all}.
Further mathematical details are provided in Appendix \ref{apx: Likelihood Ratio Test}.

The applicability of the \gls{lrt} in this context is ensured by the fact that the models in question are nested.
Specifically, the model based solely on \gls{ecg}-derived features can be viewed as a special case of the full model, obtained by constraining the coefficients of the baseline covariates to zero.
During the training phase of the full model, we utilized the same data folds that were utilized for the model using only the \gls{ecg}-derived features, ensuring that the \gls{lrt} statistics were computed consistently, providing a reliable basis for comparing the models.
When applying the \gls{lrt}, we set a significance level of $\alpha=0.05$. 
Given that the \gls{lrt} was repeated for each validation fold, we considered the multiple testing framework and adjusted the confidence level using the Bonferroni correction to $\alpha= 0.01$. 
In all scenarios, we rejected the null hypothesis of the \gls{lrt}, indicating that the full model adequately explains the \gls{ecg} dataset compared to the model based solely on \gls{ecg}-derived features. 
For completeness, we report the p-values from each of the five validation folds: for \gls{af}, we obtained p-values of 0.95, 0.98, 0.97, 0.97, and 0.96; for \gls{mi}, we obtained p-values of 0.71, 0.75, 0.69, 0.64, and 0.75; and for \gls{sbr}, we obtained p-values of 0.98, 0.97, 0.94, 0.92, and 0.91.

\section{Discussion and conclusion}\label{sec:discussion}

We explored the possibility of constructing a powerful \gls{lr}-based interpretable model to predict abnormal \gls{ecg} by feeding it with features extracted from \gls{1ddnn}-derived reconstructions.
The ultimate scope of this research was to determine whether such an interpretable model could be employed to accurately predict abnormalities in \gls{ecg}s in line with clinical experience or improve it by revealing novel interconnections captured by the \gls{1dcnn}.

In addition to this, our findings in Section \ref{sec: Comapring_ Interpretable_Predictions}, where we refined the \gls{lr} model by incorporating baseline variables like sex and age alongside \gls{ecg}-derived features, further emphasized the importance of enhancing model accuracy and clinical relevance. 
This refinement led to a noticeable, albeit slight, improvement in predictive metrics, particularly for the full model.

The rising incidence of cardiovascular disease, driven by an aging global population, has made it a leading cause of death in many countries. 
\gls{ann} models can help promptly identify arrhythmias and heart attacks, even though these models must provide interpretable predictions aligned with clinical experience to ensure reliability and transparency.

The low relative errors of \gls{3mcs} between the \gls{1dcnn} and the \gls{lr} model indicated that irregular \gls{ecg}s can be accurately classified in an interpretable manner by explaining the \gls{1dcnn} and extracting features from it. 
The \gls{lr} model demonstrated performance comparable to the \gls{1dcnn} when evaluated with metrics such as \gls{auroc}, \gls{auprc}, and \gls{mcc}.
The comparative analysis in Section \ref{sec: Comapring_ Interpretable_Predictions}, which demonstrated only slight improvements across several metrics when integrating baseline covariates, does not find any confirmation about a statistical significance among a straight difference between the two approaches, i.e., the one that combines \gls{ecg}-derived features with patient-specific information and the other that considers \gls{ecg}-derived features only.
Despite being more clinically relevant and more comprehensive, the full model reinforces the idea that the \gls{ecg}-derived features can actually explain the classification of irregular \gls{ecg}, even without the need for additional baseline covariates.

By leveraging the explanation of the pattern activity of \gls{1dcnn}, we showed it is possible to classify \gls{ecg}s in an interpretable manner. 
Post-hoc inspection of the \gls{lr} model revealed that the features extracted are grounded in clinical knowledge.
For \gls{af}, we found that the interpretable predictions are based on straight clinical signs such as the absence of the P-wave.
Similarly to this, for \gls{sbr}, we found straight clinical signs suggesting a slowing down of the heart's beat; specifically, a single normal heartbeat within a 2-second window, with no other structures or artifacts suggesting another beat before or after.
In a more complex situation such as the \gls{mi}, we observed that evaluating the presence of a mixture of different clinical signs, that do not necessarily describe the same stage of \gls{mi}, represents a strategy adopted to better generalize the classification of \gls{mi}s.
For complex cases like \gls{mi}, a mixture of clinical signs was used to generalize \gls{mi} classification, e.g., the ST segment depression in leads II, III, and avF, along with tall waves in V6, and the T-Wave inversion in leads V1-V5.

Our findings underscore the potential for creating interpretable models that can enhance clinical assessments and improve ECG classification by leveraging \gls{ann} models' computational power.
This study contributes to warning about the risks of relying solely on \gls{xai} to interpret \gls{ann} predictions.
Instead, it proposes a way to integrate \gls{xai}-derived features into more interpretable contexts with the ultimate scope of improving interpretability and accuracy in automated \gls{ecg} classification.



We have illustrated a methodology for constructing an interpretable setting for classifying irregular \gls{ecg}s within which the explanation of the \gls{1dcnn} pattern activity is exploited to derive potential novel covariates.
However, we stress that this application had the finality here to present a methodology devised for \gls{ecg}s data only.
More broad studies involving even more complex sources of clinical data are then needed to consolidate and extend this methodology. 
In addition, another open question remains on the sensitivity of the \gls{ann} methods selected as well as the \gls{xai} method adopted to extract essential features for the \gls{ecg}s classification.
As introduced in Section \ref{sec: LSTM-based predictions}, a \gls{1dcnnlstm} can also outstandingly classify \gls{ecg}s, while capturing different or similar patterns compared to the \gls{1dcnn} 
However, the \gls{1ddnn} can be employed to explain \gls{1dcnn}s only; 
an explanation thought a different method could affect the covariate derivation during the construction of the desired interpretable modelling framework.

In conclusion, we shed some light on the importance of integrating interpretability into advanced \gls{ann}s for \gls{ecg} classification, providing a valuable bridge between computational power and clinical applicability. 
By ensuring that predictions are not only accurate but also grounded in well-founded clinical principles, our methodology aims to enhance the reliability and transparency of predictions.

\begin{appendices}
\noappendicestocpagenum
\addappheadtotoc

\section{Model Selection Details}

\subsection{Hyperparameter Tuning of 1-D CNN}\label{apx: hyperparamenters}
As introduced in Section \ref{sec:model_fit}, the configuration for the \gls{1dcnn} of each of the three classification tasks was selected using the grid-search method, that is, the \gls{1dcnn} performance was explored by considering a combination of hyperparameter distributed along a fine grid of values.
While searching for the optimal networks' hyperparameters, we utilized a 5-fold cross-validation process for each 

The model's selection was then drawn by considering the model with the highest predictive power.
In Section \ref{sec:model_fit}, we introduced that the evaluation of the overall predictive power was then assessed by considering 3 different metrics, i.e., \gls{auroc}, \gls{auprc}, and \gls{mcc}.
As discussed, each of those metrics reflects a classifier's prediction under different aspects; so, for the sake of completeness, we opted for analysing all these metrics at the same time.
Briefly, in a binary classification task, \gls{auroc} estimates the probability that binary classifiers guess correctly the label of a target instance.
However, such a metric is sensitive to the \emph{sensitivity}, and it might lead to optimistic results when a few positive instances are included in the dataset.
The \gls{auprc} is the area under the curve \emph{precision} vs. \emph{recall}.
We recall that the precision is the ratio proportion of true positive predictions among all positive predictions, while the recall is the ratio proportion of true positive predictions among all positive predictions.
The \gls{auprc} can help obtain more information about the ability of the model to distinguish true and false positives, especially in the context of imbalanced datasets where the precision-recall trade-off is more informative than accuracy-based metrics.
Similarly, the \gls{mcc} provides insight into the correct level of correlation between predicted and actual labels, countering overly optimistic conclusions that \gls{auroc} might suggest. 
It offers a more comprehensive and intuitive assessment of the model's predictions.
In light of all these considerations, we then considered a metric derived from the product of all these three metrics; we denote it as \gls{3mcs}.
The \gls{3mcs} was defined thought the formula 
\begin{equation}
    \operatorname{3MCS}(\tilde{y}, y)  = \operatorname{AUROC}(\tilde{y}, y)\operatorname{AURPC}(\tilde{y}, y) \frac{1+\operatorname{MCC}(\tilde{y}, y)}{2};
\end{equation}
with $\tilde{y}$ and $y$ some generic pairs of prediction and ground labels, respectively.

A detailed report of all metrics, for all configurations considered, is shown, for \gls{af}, in Table \ref{tab: Model's Selection AF}; for \gls{mi}, in Table \ref{tab: Model's Selection MI}; for \gls{sbr}, in Table \ref{tab: Model's Selection SB}. 
\begin{table}[]
    \centering
\begin{tabular}{rrrllll}
\toprule
 Filters &  Kernal\_Size &  Deepness &       AUC  &     AUPRC &       MCC &      3MCS  \\
\midrule
       8 &            3 &         2 & 0.87±0.01 & 0.93±0.01 & 0.56±0.02 & 0.63±0.03 \\
       8 &            3 &         3 & 0.78±0.01 & 0.89±0.01 & 0.43±0.02 & 0.49±0.01 \\
       8 &            3 &         4 & 0.82±0.02 & 0.91±0.01 & 0.47±0.03 & 0.54±0.03 \\
       8 &            5 &         2 & 0.89±0.01 & 0.94±0.01 & 0.61±0.03 & 0.45±0.03 \\
       8 &            5 &         3 & 0.83±0.01 & 0.92±0.01 & 0.48±0.03 & 0.57±0.03 \\
       8 &            5 &         4 & 0.86±0.01 & 0.94±0.01 & 0.51±0.01 & 0.61±0.01 \\
       8 &            9 &         2 & 0.84±0.01 & 0.92±0.01 & 0.48±0.02 & 0.57±0.01 \\
       8 &            9 &         3 & 0.79±0.01 & 0.89±0.01 & 0.42±0.01 & 0.49±0.01 \\
       8 &            9 &         4 & 0.82±0.01 & 0.90±0.01 & 0.47±0.01 & 0.54±0.01 \\
      16 &            3 &         2 & 0.94±0.01 & 0.97±0.01 & 0.77±0.02 & 0.80±0.02 \\
      16 &            3 &         3 & 0.91±0.01 & 0.95±0.01 & 0.70±0.01 & 0.74±0.01 \\
      16 &            3 &         4 & 0.90±0.01 & 0.95±0.01 & 0.60±0.02 & 0.68±0.02 \\
      16 &            5 &         2 & 0.94±0.01 & 0.97±0.01 & 0.75±0.01 & 0.78±0.01 \\
      16 &            5 &         3 & 0.90±0.01 & 0.95±0.01 & 0.63±0.01 & 0.70±0.02 \\
      16 &            5 &         4 & 0.90±0.01 & 0.94±0.01 & 0.62±0.02 & 0.69±0.02 \\
      16 &            9 &         2 & 0.90±0.01 & 0.95±0.01 & 0.66±0.01 & 0.72±0.02 \\
      16 &            9 &         3 & 0.88±0.01 & 0.94±0.01 & 0.60±0.03 & 0.66±0.03 \\
      16 &            9 &         4 & 0.86±0.01 & 0.92±0.01 & 0.54±0.02 & 0.61±0.02 \\
      32 &            3 &         2 & 0.93±0.01 & 0.95±0.01 & 0.81±0.01 & 0.79±0.02 \\
      32 &            3 &         3 & 0.94±0.01 & 0.97±0.01 & 0.76±0.02 & 0.79±0.02 \\
      32 &            3 &         4 & 0.94±0.01 & 0.97±0.01 & 0.72±0.01 & 0.67±0.02 \\
      32 &            5 &         2 & 0.94±0.01 & 0.96±0.01 & 0.80±0.01 & 0.81±0.01 \\
      32 &            5 &         3 & 0.94±0.01 & 0.97±0.01 & 0.73±0.02 & 0.79±0.03 \\
      32 &            5 &         4 & 0.95±0.01 & 0.98±0.01 & 0.78±0.02 & 0.83±0.02 \\
      32 &            9 &         2 & 0.94±0.01 & 0.96±0.01 & 0.78±0.01 & 0.81±0.01 \\
      32 &            9 &         3 & 0.94±0.01 & 0.97±0.01 & 0.76±0.01 & 0.79±0.01 \\
      32 &            9 &         4 & 0.95±0.01 & 0.97±0.01 & 0.76±0.02 & 0.80±0.02 \\
\bottomrule
\end{tabular}
    \caption{Grid Search selection for the \gls{af} datasets. Columns with the hyperparameters investigated and the metrics' values are shown.
    The values refer to the validation phase.}
    \label{tab: Model's Selection AF}
\end{table}

\begin{table}[]
    \centering
    \begin{tabular}{rrrllll}
\toprule
 Filters &  Kernel Size &  Deepness &       AUC &     AUPRC &       MCC &      3MCS \\
\midrule
       8 &            3 &         2 &  1.0±0.01 &  1.0±0.01 & 0.99±0.01 & 0.99±0.01 \\
       8 &            3 &         3 &  1.0±0.01 & 0.99±0.01 & 0.98±0.01 & 0.98±0.01 \\
       8 &            3 &         4 & 0.97±0.01 & 0.97±0.01 & 0.82±0.01 & 0.85±0.01 \\
       8 &            5 &         2 &  1.0±0.01 &  1.0±0.01 & 0.99±0.01 & 0.99±0.01 \\
       8 &            5 &         3 & 0.99±0.01 & 0.99±0.01 & 0.94±0.01 & 0.95±0.01 \\
       8 &            5 &         4 & 0.98±0.01 & 0.98±0.01 & 0.88±0.02 & 0.91±0.01 \\
       8 &            9 &         2 &  1.0±0.01 &  1.0±0.01 & 0.99±0.01 & 0.99±0.01 \\
       8 &            9 &         3 & 0.99±0.01 & 0.99±0.01 & 0.94±0.01 & 0.95±0.01 \\
       8 &            9 &         4 & 0.96±0.01 & 0.97±0.01 & 0.78±0.01 & 0.82±0.01 \\
      16 &            3 &         2 &  1.0±0.01 &  1.0±0.01 & 0.99±0.01 & 0.99±0.01 \\
      16 &            3 &         3 &  1.0±0.01 & 0.99±0.01 & 0.99±0.01 & 0.99±0.01 \\
      16 &            3 &         4 & 0.98±0.01 & 0.98±0.01 & 0.91±0.01 & 0.92±0.01 \\
      16 &            5 &         2 &  1.0±0.01 &  1.0±0.01 & 0.99±0.01 & 0.99±0.01 \\
      16 &            5 &         3 &  1.0±0.01 & 0.99±0.01 & 0.99±0.01 & 0.99±0.01 \\
      16 &            5 &         4 & 0.99±0.01 & 0.99±0.01 & 0.96±0.01 & 0.97±0.01 \\
      16 &            9 &         2 &  1.0±0.01 &  1.0±0.01 & 0.99±0.01 & 0.99±0.01 \\
      16 &            9 &         3 &  1.0±0.01 &  1.0±0.01 & 0.99±0.01 & 0.99±0.01 \\
      16 &            9 &         4 & 0.99±0.01 & 0.99±0.01 & 0.93±0.01 & 0.94±0.01 \\
      32 &            3 &         2 &  1.0±0.01 &  1.0±0.01 & 0.99±0.01 & 0.99±0.01 \\
      32 &            3 &         3 &  1.0±0.01 &  1.0±0.01 & 0.99±0.01 & 0.99±0.01 \\
      32 &            3 &         4 &  1.0±0.01 & 0.99±0.01 & 0.99±0.01 & 0.98±0.01 \\
      32 &            5 &         2 &  1.0±0.01 &  1.0±0.01 & 0.99±0.01 & 0.99±0.01 \\
      32 &            5 &         3 &  1.0±0.01 &  1.0±0.01 & 0.99±0.01 &  1.0±0.01 \\
      32 &            5 &         4 &  1.0±0.01 &  1.0±0.01 & 0.98±0.01 & 0.99±0.01 \\
      32 &            9 &         2 &  1.0±0.01 &  1.0±0.01 &  1.0±0.01 &  1.0±0.01 \\
      32 &            9 &         3 &  1.0±0.01 &  1.0±0.01 &  1.0±0.01 &  1.0±0.01 \\
      32 &            9 &         4 &  1.0±0.01 &  1.0±0.01 & 0.98±0.01 & 0.98±0.01 \\
\bottomrule
\end{tabular}
    \caption{Grid Search selection for the \gls{mi} datasets. Columns with the hyperparameters investigated and the metrics' values are shown.
    The values refer to the validation phase.}
    \label{tab: Model's Selection MI}
\end{table}

\begin{table}[]
    \centering
    \begin{tabular}{rrrllll}
\toprule
 Filters &  Kernal\_Size &  Deepness &       AUC &     AUPRC &       MCC &      3MCS \\
\midrule
       8 &            3 &         2 & 0.73±0.01 & 0.59±0.01 & 0.32±0.01 & 0.59±0.01 \\
       8 &            3 &         3 & 0.77±0.01 & 0.62±0.01 & 0.37±0.01 & 0.61±0.01 \\
       8 &            3 &         4 & 0.87±0.01 & 0.79±0.01 & 0.58±0.01 & 0.73±0.01 \\
       8 &            5 &         2 & 0.84±0.01 & 0.75±0.01 & 0.51±0.01 & 0.73±0.01 \\
       8 &            5 &         3 & 0.79±0.01 & 0.65±0.01 & 0.43±0.01 & 0.61±0.01 \\
       8 &            5 &         4 & 0.94±0.01 & 0.88±0.01 & 0.74±0.01 & 0.84±0.01 \\
       8 &            9 &         2 & 0.75±0.01 & 0.63±0.01 & 0.35±0.01 & 0.62±0.01 \\
       8 &            9 &         3 & 0.92±0.01 & 0.86±0.01 & 0.68±0.01 & 0.82±0.01 \\
       8 &            9 &         4 & 0.91±0.01 & 0.84±0.01 & 0.66±0.01 & 0.79±0.01 \\
      16 &            3 &         2 &  0.8±0.01 &  0.7±0.01 & 0.44±0.01 & 0.65±0.01 \\
      16 &            3 &         3 & 0.88±0.01 &  0.8±0.01 & 0.60±0.01 & 0.75±0.01 \\
      16 &            3 &         4 & 0.89±0.01 & 0.81±0.01 & 0.61±0.01 & 0.75±0.01 \\
      16 &            5 &         2 & 0.86±0.01 & 0.77±0.01 & 0.55±0.01 & 0.72±0.01 \\
      16 &            5 &         3 & 0.82±0.01 &  0.7±0.01 & 0.48±0.01 & 0.70±0.01 \\
      16 &            5 &         4 & 0.83±0.01 & 0.69±0.01 & 0.50±0.01 & 0.64±0.01 \\
      16 &            9 &         2 & 0.87±0.01 & 0.79±0.01 & 0.58±0.01 & 0.73±0.01 \\
      16 &            9 &         3 & 0.84±0.01 & 0.72±0.01 & 0.52±0.01 & 0.67±0.01 \\
      16 &            9 &         4 & 0.96±0.01 & 0.93±0.01 & 0.79±0.01 & 0.90±0.01 \\
      32 &            3 &         2 & 0.86±0.01 & 0.77±0.01 & 0.55±0.01 & 0.71±0.01 \\
      32 &            3 &         3 & 0.86±0.01 & 0.76±0.01 & 0.56±0.01 & 0.71±0.01 \\
      32 &            3 &         4 & 0.85±0.01 & 0.74±0.01 & 0.55±0.01 & 0.67±0.01 \\
      32 &            5 &         2 & 0.89±0.01 & 0.81±0.01 & 0.61±0.01 & 0.76±0.01 \\
      32 &            5 &         3 & 0.87±0.01 & 0.78±0.01 & 0.58±0.01 & 0.73±0.01 \\
      32 &            5 &         4 & 0.97±0.01 & 0.94±0.01 & 0.83±0.01 & 0.92±0.01 \\
      32 &            9 &         2 & 0.90±0.01 & 0.82±0.01 & 0.64±0.01 & 0.60±0.01 \\
      32 &            9 &         3 & 0.97±0.01 & 0.95±0.01 & 0.84±0.01 & 0.93±0.01 \\
      32 &            9 &         4 & 0.98±0.01 & 0.94±0.01 & 0.86±0.01 & 0.93±0.01 \\
\bottomrule
\end{tabular}
\caption{Grid Search selection for the \gls{sbr} datasets. Columns with the hyperparameters investigated and the metrics' values are shown.
The values refer to the validation phase.}
    \label{tab: Model's Selection SB}
\end{table}

\subsection{Hyperparameter Tuning of LSTM-based models}\label{apx: model's selection}

Along with the \gls{1dcnn} model, \gls{lstm}-based models, including the standalone \gls{lstm} and the hybrid \gls{1dcnnlstm}, were also considered during the model selection process.

When considering a \gls{lstm} model we referred to one \gls{lstm} layer that is fed with \gls{ecg}s and outputs a latent description of \gls{ecg}s with dimensionality equal to the number of units of the \gls{lstm} layer.
Such a latent description is then propagated through a dense layer with a single output node with a sigmoid activation function.
This dense layer provides the final prediction.
The term \gls{1dcnnlstm} model refers to an architecture based on the \gls{1dcnn}, as introduced in Section \ref{sec:model_fit}, where the Flattern layer is then replaced by a \gls{lstm} layer. 
Both models were embedded with the same optimizer and loss function utilized for the \gls{1dcnn}.

In our search for the best configurations for these two models, we chose to explore a single hyperparameter: the number of \gls{lstm} units. For the \gls{1dcnnlstm} model, this choice reflects the potential impact that the \gls{lstm} layer could have on the performance of the \gls{1dcnn} architecture, as discussed in Section \ref{apx: hyperparamenters}. 
As a result, the hyperparameters related to the convolutional layers of the \gls{1dcnnlstm} were fixed to the optimal values outlined in Section \ref{apx: hyperparamenters}. 
For the \gls{lstm} model, the optimal \gls{lstm} units are shown in Table \ref{tab: lstm_metrics}, while for the \gls{1dcnnlstm} model, see \ref{tab: 1dcnn_lstm_metrics}.
\begin{table}[]
    \centering
    \begin{tabular}{rllll}
    \toprule
    & & \textbf{AF dataset}& & \\
    \midrule
    LSTM units &       AUC &     AUPRC &       MCC &      3MCS \\
    \midrule
          2 & 0.74±0.07 & 0.85±0.04 & 0.38±0.12 &  0.46±0.1 \\
          4 & 0.72±0.05 & 0.85±0.03 & 0.28±0.12 & 0.42±0.08 \\
          8 & 0.64±0.07 & 0.79±0.04 & 0.18±0.11 & 0.32±0.09 \\
         16 & 0.68±0.06 & 0.83±0.03 &  0.2±0.12 & 0.36±0.08 \\
         32 &  0.8±0.09 &  0.9±0.05 & 0.47±0.19 & 0.59±0.15 \\
         64 & 0.64±0.07 &  0.8±0.04 & 0.19±0.13 &  0.33±0.1 \\
    \midrule
    & & \textbf{MI dataset}& & \\
    \midrule
      LSTM units &       AUC &     AUPRC &       MCC &      3MCS \\
    \midrule
          2 & 0.57±0.07 & 0.67±0.06 &  0.1±0.11 & 0.24±0.09 \\
          4 & 0.66±0.06 & 0.72±0.05 & 0.24±0.07 & 0.31±0.07 \\
          8 & 0.69±0.05 & 0.73±0.04 & 0.35±0.08 & 0.36±0.08 \\
         16 & 0.73±0.07 & 0.77±0.06 & 0.41±0.11 & 0.43±0.12 \\
         32 & 0.76±0.06 &  0.8±0.04 & 0.43±0.11 &  0.46±0.1 \\
         64 & 0.86±0.05 & 0.88±0.04 & 0.59±0.11 &  0.63±0.1 \\
    \midrule
    & & \textbf{SB dataset}& & \\
    \midrule
      LSTM units &       AUC &     AUPRC &       MCC &      3MCS \\
    \midrule
              2 & 0.62±0.05 & 0.48±0.05 & 0.12±0.09 & 0.18±0.05 \\
          4 & 0.58±0.02 & 0.46±0.02 & 0.13±0.02 & 0.15±0.01 \\
          8 &  0.71±0.1 & 0.63±0.12 &  0.38±0.2 & 0.41±0.18 \\
         16 &  0.6±0.03 & 0.46±0.01 & 0.14±0.02 & 0.16±0.01 \\
         32 & 0.56±0.01 & 0.44±0.01 & 0.11±0.02 & 0.14±0.01 \\
         64 & 0.64±0.03 &  0.5±0.02 & 0.21±0.03 &  0.2±0.02 \\
    \bottomrule
    \end{tabular}
    \caption{Validation Metrics for different values of the \gls{lstm} units of the \gls{lstm} model }
    \label{tab: lstm_metrics}
\end{table}

\begin{table}[]
    \centering
    \begin{tabular}{rllll}
    \toprule
    & & \textbf{AF dataset}& & \\
    \midrule
    LSTM units &       AUC &     AUPRC &       MCC &      3MCS \\
    \midrule
          2 & 0.95±0.01 & 0.97±0.01 & 0.77±0.01 & 0.82±0.01 \\
          4 & 0.69±0.07 & 0.85±0.04 & 0.17±0.16 & 0.38±0.12 \\
          8 & 0.86±0.08 & 0.94±0.03 & 0.51±0.21 & 0.65±0.15 \\
         16 & 0.97±0.01 & 0.99±0.01 & 0.75±0.04 &  0.72±0.03 \\
         32 & 0.92±0.07 & 0.96±0.04 &  0.76±0.2 & 0.67±0.15 \\
         64 &  0.7±0.01 & 0.86±0.01 & 0.05±0.05 & 0.32±0.02 \\
    \midrule
    & & \textbf{MI dataset}& & \\
    \midrule
      LSTM units &       AUC &     AUPRC &       MCC &      3MCS \\
    \midrule
          2 & 0.89±0.04 & 0.92±0.03 & 0.66±0.09 & 0.69±0.08 \\
          4 & 0.92±0.02 & 0.95±0.01 & 0.74±0.06 & 0.77±0.06 \\
          8 & 0.96±0.01 & 0.97±0.01 & 0.85±0.04 & 0.87±0.04 \\
         16 & 0.94±0.04 & 0.96±0.02 &  0.78±0.1 & 0.82±0.09 \\
         32 & 0.99±0.01 & 0.99±0.01 & 0.92±0.01 & 0.95±0.01 \\
         64 & 0.99±0.01 & 0.99±0.01 & 0.93±0.02 & 0.95±0.01 \\
    \midrule
    & & \textbf{SB dataset}& & \\
    \midrule
    \toprule
    LSTM units &      AUC &    AUPRC &       MCC &      3MCS \\
    \midrule
          2 & 1.0±0.01 & 1.0±0.01 & 0.98±0.01 & 0.99±0.01 \\
          4 & 1.0±0.01 & 1.0±0.01 & 0.97±0.01 & 0.98±0.01 \\
          8 & 1.0±0.01 & 1.0±0.01 & 0.98±0.01 & 0.98±0.01 \\
         16 & 1.0±0.01 & 1.0±0.01 & 0.97±0.01 & 0.98±0.01 \\
         32 & 1.0±0.01 & 1.0±0.01 & 0.98±0.01 & 0.98±0.01 \\
         64 & 1.0±0.01 & 1.0±0.01 & 0.98±0.01 & 0.99±0.01 \\
    \bottomrule
    \end{tabular}
    \caption{Validation Metrics for different values of the \gls{lstm} units of the \gls{1dcnnlstm} }
    \label{tab: 1dcnn_lstm_metrics}
\end{table}

\section{Deconvolutional Neural Networks}

In this section, we provide further technical details about the \gls{1ddnn} employed in the manuscript. 
As previously discussed, the key characteristic of the \gls{1ddnn} is its ability to invert the hidden layers of a \gls{1dcnn} model, allowing us to analyze the information captured during the learning phase \cite[]{zeiler2010deconvolutional}. 
The primary function of the \gls{1ddnn} is to reverse the deepest feature map of the \gls{1dcnn} back to the input space. 
Despite its layered structure, similar to the \gls{1dcnn}, the \gls{1ddnn} does not undergo a learning phase. 
Instead, it consists of a sequence of non-trainable layers designed to replicate the inversion of the \gls{1dcnn} hidden layers. 

We recall that a \gls{1dcnn} are mainly based on the sequential repetition of layers such as \emph{Convolutional Layer} $\to$ \emph{Activation Layer} $\to$ \emph{Maxpooling Layer}.
Accordingly, the \gls{dnn}'s layers are designed to invert such a sequence; for example utilizing the following other sequences of layers: \emph{Unpooling Layer} $\to$ \emph{Deactivation Layer} $\to$ \emph{Transposed Convolutional Layer}.

We examine and comment now all these layers.
As known, pooling operators (such as the max-pooling or average-pooling layers) are non-linear operators that do not admit the inverse operator.
This problem precludes us from the possibility of obtaining the exact reconstruction of the input of any pooling layer from its output.
Approximations, however, represent a good compromise to obtain a robust reconstruction without a crucial loss of information; in this case, spline interpolation is of very broad use.
In the last decade years, when dealing with the max-pooling layers, the inversion thought the method of spline interpolation was abandoned in favour of the \emph{switch variables} proposed by \cite{Zeiler2014}.
When reconstructing the input of a max-pooling layer, the switch variables rearrange the pooled values in such a way that they are relocated to their original locations. 
This fact ensures a better reconstruction with much less aliasing or digital artefacts.
However, the method of switching variables does not recommend any strategy regarding how to treat the other values, namely those values that are discarded by the pooling layer. 
No suggestions are given; they can be imputed through a spline or any other imputation method, depending on the situation.

In Section \ref{sec:model_fit}, we established that the semi-orthogonality constraint of convolutional filters is essential for ensuring that convolutional and transposed convolutional layers are inverses of each other. The following discussion will derive how this constraint is necessary for making convolutional inversion more feasible.
We shall make use of \emph{Einstein's notation}; namely, repeated indexes replicate a summation over a of indexed terms in a formula. 
This choice will help us to have an uncluttered notation.
For the sake of clarity, summations will be written explicitly when needed.

\subsection{Exact inversion of 1-D convolutional layers }\label{apx: inversion_1d_convolutional_layer}

Let us consider the case of a 1-D convolutional layer.
With denote with $w_{ijk}$ a convolutional kernels, where the indexes $i,$$j,$and $k$ denotes, respectively,
the output feature map, its temporal location, and its input feature map.
The symbol $X_{jk}$ refers to a generic instance of time-series data, where the first index indicates the temporal location and the second index denotes a feature of the time series.
To avoid excessive notation, we assume that all indices range over a finite set of consecutive integers.

Without loss of generality, we will consider a 1-D convolutional layer that is directly connected to and follows the input layer. 
As known, the \emph{forward propoagation} through this layer is ruled by the convolutional operator, so it reads:
\begin{equation}\label{eq:1d_fwd_conv}
    \Gamma_{i,j} = w_{i,j-r,k}X_{k, r};
\end{equation}
where $\Gamma_{i, j}$ is the $i-th$ feature map at the temporal location $j.$

We now seek to invert $\Gamma_{i, j}$ to retrieve $X_{ij}$.
To achieve this, we will use a \emph{backward propagation} of $\Gamma_{i, j}$ through a 1-D Transposed convolutional layer.
We assume that the 1-D Transposed convolutional layer utilizes the same weights of \eqref{eq:1d_fwd_conv}.
So we have
\begin{equation}
    \Upsilon_{i,j} =  w^{T}_{i, j-r, k}\Gamma_{k,r};
\end{equation}
or equivalently
\begin{equation}\label{eq:1d_bwd_conv}
    \Upsilon_{i,j} =  w_{i, j+r, k}\Gamma_{k,r}.
\end{equation}
When the output of the backward propagation coincides with the input in the forward propagation, we have that \eqref{eq:1d_fwd_conv} and \eqref{eq:1d_bwd_conv} are each other's inverse operator.
Thus, we write
\begin{equation}
    X_{i, j} = w_{i, r, k}w_{k, s, l}X_{l, j+r-s}. 
\end{equation}
The last equation is valid if 
\begin{equation}
    w_{i, r, k}w_{k, s, l} = \delta_{il}\delta_{sr}; 
\end{equation}
or equivalently 
\begin{equation}\label{eq:discerete_orthogonality_condition_conv_1d}
    w_{i, r, k}w_{k, r, l} = \delta_{il}; 
\end{equation}
with $\delta$ denoting a Kronecker's delta.
The \eqref{eq:discerete_orthogonality_condition_conv_1d} is the orthogonality constraint on filters.
Notice that the term orthogonality is correct as long as the number of feature maps is equal to the number of features in the input data. 
When this is not the case, it is more appropriate to speak of \emph{semi-orthogonality} constraint.

\subsection{2-D Convolutional Layer}\label{apx: inversion_2d_convolutional_layer}
The results obtained for 1-D convolutions can be readily extended to a 2-D convolutional layer. 
In this context, we define $ w_{i,r,s,l} $ as a generic 2-D convolutional filter, where the middle indices represent the filter's width and length, while the first and last indices denote the output and input features, respectively. 
We now denote a 2-D image as $X_{ijk}$, where $j$ and $k$ indicate the spatial locations within the image.

Following the same argumentation of \ref{apx: inversion_1d_convolutional_layer}, the 2-D forward propagation rule reads
\begin{equation}
    \Gamma_{i,j,k} = w_{i, r, s, l}X_{l, j-r, k-s};
\end{equation}
while the backward propagation rule (through a 2-D transposed convolutional layer) is 
\begin{equation}
    \Upsilon_{i,j,k} = w_{i,r, s, l}\Gamma_{l, j+r, k+s}.
\end{equation}
To ensure the inversion of the convolutional layer, we impose the condition that the input of the forward propagation must be equal to the output of the backward propagation.
Thus, we have
\begin{equation}
    X_{i,j,k} = w^{T}_{i,r, s, l}w_{i,p, q, m}X_{m, j+r-q, k+s-q}.
\end{equation}
The equality between the two sides of the last equation is achieved as
\begin{equation}
    w_{i,r, s, l}w_{l,p, q, m} = \delta_{im}\delta_{r,p}\delta_{s,q}.
\end{equation}
or simply 
\begin{equation}\label{eq:discerete_orthogonality_condition_conv_2d}
    w_{i, r, s, k}w_{k, r, s, l} = \delta_{il}. 
\end{equation}
The \eqref{eq:discerete_orthogonality_condition_conv_2d} is the semi-orthogonality constraint ensuring 2-D transposed convolutional layers are the inverse layer of 2-D convolutional layer; and vice versa.

\subsection{Dense Layer}\label{apx: inversion_dense_layer}
At last, we shall consider exact inversion for a Dense Layer.
Although a Dense Layer is not ruled by convolutions, we would like to show how the orthogonality constraints are crucial to invert the linear product involved in such a layer.
Thus, we indicate with $w_{kj}$ the weights ruling the \emph{forward propagation} from the $j$-th input through the $k$-th output.
As known the forward rule is
\begin{equation}
    \Gamma_{k} = w_{kj}X_{j}; 
\end{equation}
with $\Gamma_{k}$ the $k$-th output and $X_j$ the $j$-th input.
The \emph{backward propagation} is obtained by means of the transposed dense layer, namely
\begin{equation}
    \Upsilon_{k} = w^{T}_{kj}\Gamma_{j};
\end{equation}
or equivalently
\begin{equation}
    \Upsilon_{k} = w_{kj}\Gamma_{j}.
\end{equation}
To ensure that forward and backward propagation rules are inverse to each other we impose the condition
\begin{equation}
    X_{k} = w_{jk}w_{jl}X_{l}.
\end{equation}
The last equation is satisfied if and only if
\begin{equation}\label{eq:orthogonality_condition_dense}
    w_{jk}w_{jl} = \delta_{kl}.
\end{equation}
The \eqref{eq:orthogonality_condition_dense} represents the semi-orthogonality constraint on the kernel of the dense layer.
Hence, we proved that the inverse layer of a dense layer is the transposed dense layer, provided that the semi-orthogonality constraint on the weights $w_{jk}$ is met.

\section{Chi-squared Saliency Map}\label{apx: Chi-squared Saliency Map}

In this section, we delve deeper into the mathematical foundations of the construction of the \gls{cssm}. Using the notation established in Section \ref{sec: 1-D DNN-based saliency maps}, we denote a generic time series instance as \( X \) and its reconstruction via the \gls{1ddnn} as \( X' \).
Given that \( X' \) is a reconstruction of \( X \), it is reasonable to expect that \( X' \) retains a significant amount of information from \( X \). Therefore, for a specific timestamp \( t \), we can model the relationship as follows:

\begin{equation}
X'_t = X_t + \zeta_t,
\end{equation}
where \( \zeta_t \sim \mathcal{N}(0,\,\sigma^{2}) \). 
Rearranging this equation, we obtain:
\begin{equation}\label{eq: standardization_discrepancy_sm}
Z^{2}_{t} = \left(\frac{X'_t - X_t}{\sigma}\right)^{2},
\end{equation}
where \( Z_{t} \sim \mathcal{N}(0, 1) \) implies that \( Z^{2}_{t} \sim \chi^{2}_{1} \).
Thus, the individual observations of \( Z_t \) are distributed according to the Chi-squared distribution with one degree of freedom. The survival function denoted as \( \mathbf{P}(\chi^2_{1} > Z^2) \), provides a means to evaluate the likelihood that the discrepancies between the original data and its reconstruction are attributable to random variation rather than indicative of a significant loss of relevant information.
For the purposes of this study, this insight allows us to leverage the survival function as a \gls{sm}, indicating which salient patterns have been captured by the \gls{1dcnn} during the reconstruction process.
We express this relationship mathematically as follows:

\begin{equation}
\mathbf{P}(\chi^2_{1} > Z^2) = 1 - \frac{1}{\Gamma(1/2)} \gamma\left(1/2, \frac{Z^2}{2}\right),
\end{equation}
where:
\begin{equation}
\Gamma(k) = \int_{0}^{+\infty} d\xi \, \xi^k \exp(-\xi) \quad (\xi > 0);
\end{equation}
and 
\begin{equation}
\gamma(k, Z) = \int_{Z}^{+\infty} d\xi \, \xi^k \exp(-\xi) \quad (\xi > 0 \, \text{and} \, Z \ge 0).
\end{equation}

This equation completes the formulation of the \gls{sm} that we proposed in Section \ref{sec: 1-D DNN-based saliency maps}.

It is important to note that in Equation \eqref{eq: standardization_discrepancy_sm}, the standardization of the discrepancy \( X_t - X'_t \) was carried out under the assumption that \( \sigma \) is known. In practice, this is rarely the case. However, we can estimate \( \sigma \) using the following unbiased and consistent estimator:

\begin{equation}
\hat{\sigma} = \sqrt{\frac{1}{n-1}\sum_{k=1}^{n} (X_{k} - X'_{k})^2}.
\end{equation}

This estimator enables us to effectively approximate the variability of the discrepancies, thus facilitating the application of our \gls{sm} construction.

\section{Sensitivity analysis for saliency map-based feature extraction}\label{apx: sensitivity analysis feature extraction}
In section \ref{sec: feature extraction}, we presented our methodology to construct a set of new explainable features based on the \gls{1ddnn}-derived \gls{sm}s.
We utilized \gls{sm}s to identify and extract segments with the most salient patterns in each \gls{tsi}. 
These patterns, defined by the highest averaged saliency, help focus on the chunks of data most aligned with the model's predictions. 
Since each \gls{sm}s provides unique information for a \gls{tsi}, we employed the \gls{ks} clustering method to generalize these patterns across different \gls{tsi}s. 
The centroids from \gls{ks}, representing the most salient patterns, were used as features by evaluating their presence within each \gls{tsi} through a sliding window. 
To enhance feature separability, we applied kernel principal component analysis (KPCA) with a radial basis function kernel, with a number of KPCA components equal to that of the extracted features.

\begin{figure}
    \centering
    \includegraphics[width= .9\textwidth]{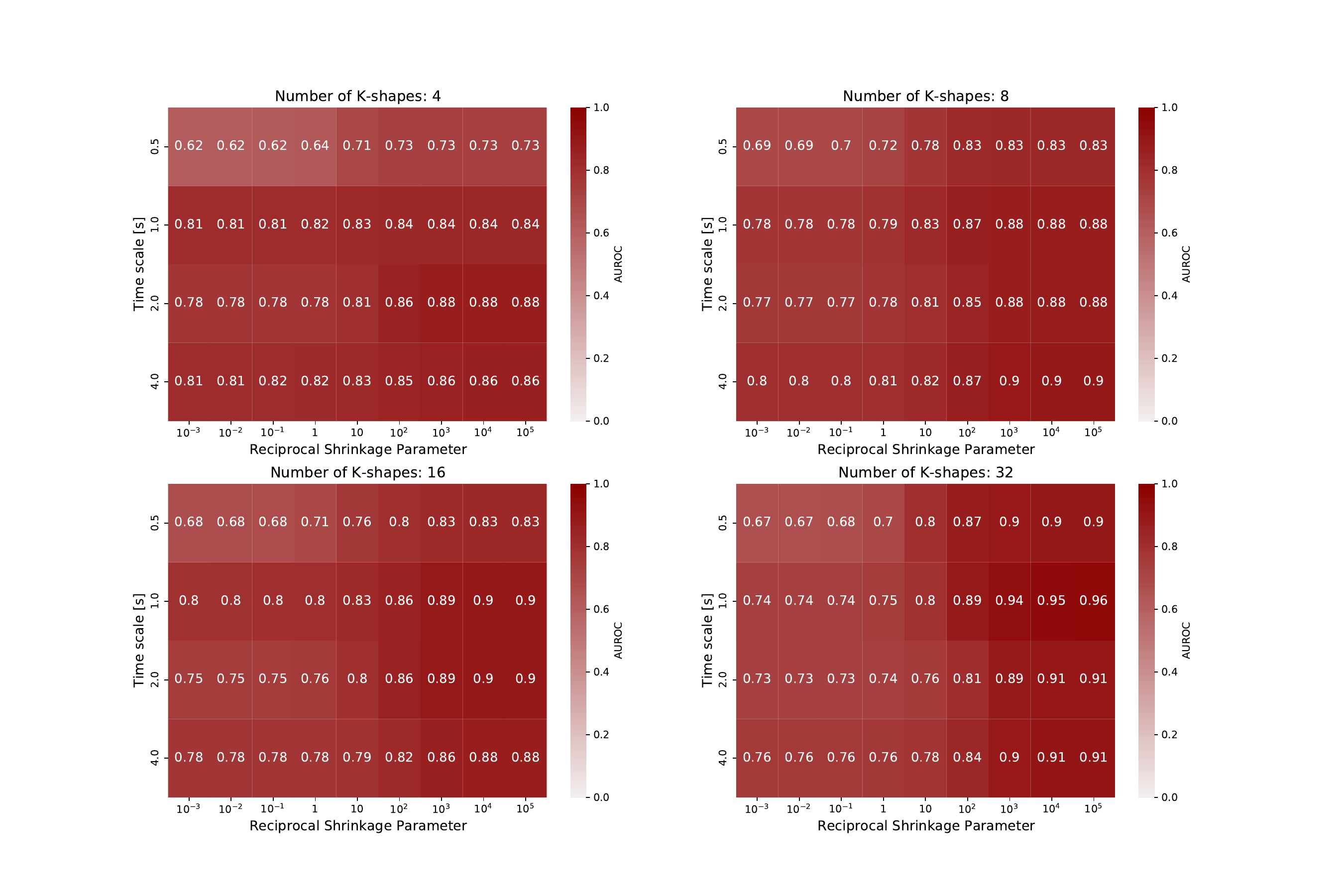}
    \caption{Sensitivity analysis for \gls{af}. Metric \gls{auroc}}
    \label{fig:sens_anal_af_auroc}
\end{figure}

\begin{figure}
    \centering
    \includegraphics[width= .9\textwidth]{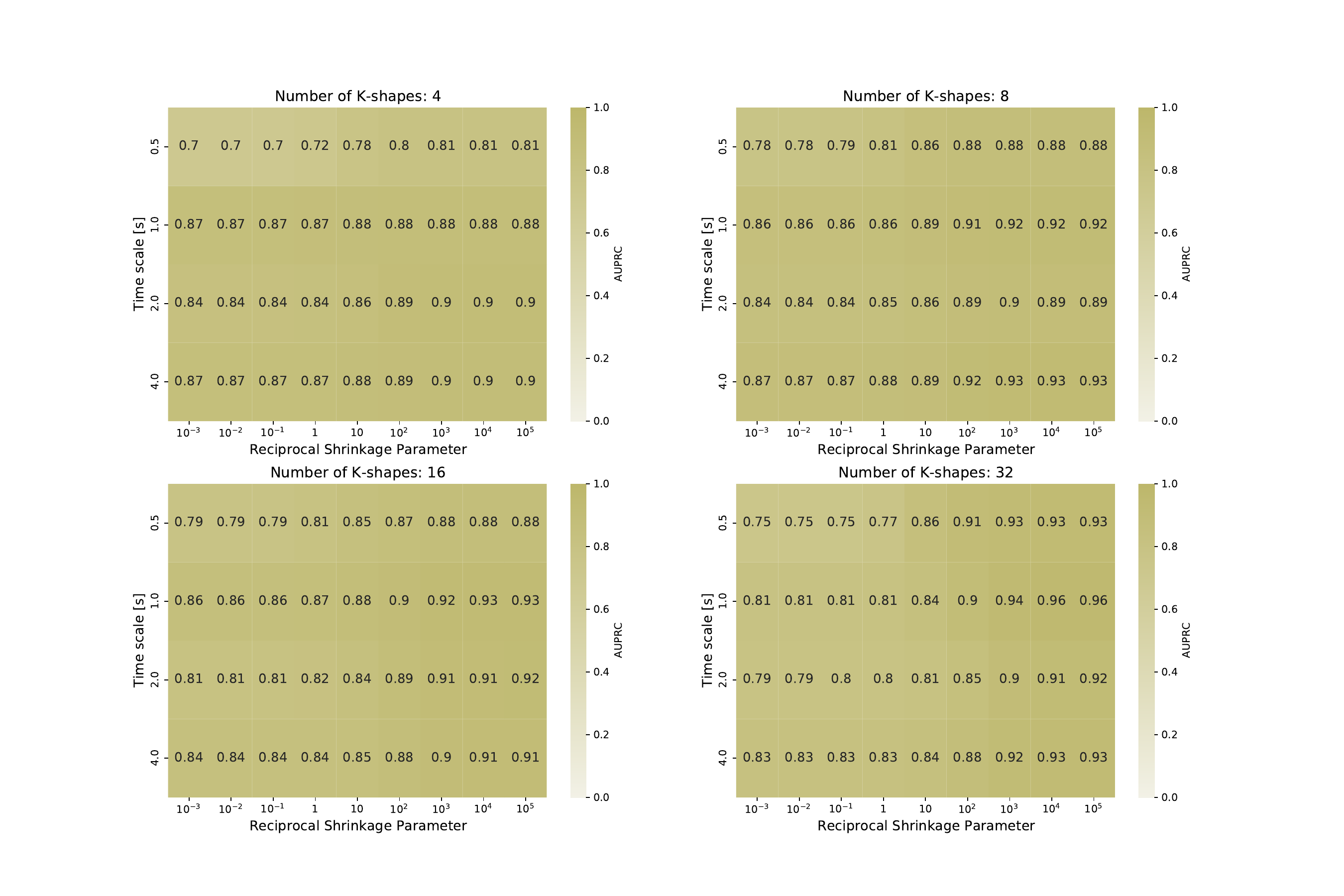}
    \caption{Sensitivity analysis for \gls{af}. Metric \gls{auprc}}
    \label{fig:sens_anal_af_auprc}
\end{figure}

\begin{figure}
    \centering
    \includegraphics[width= .9\textwidth]{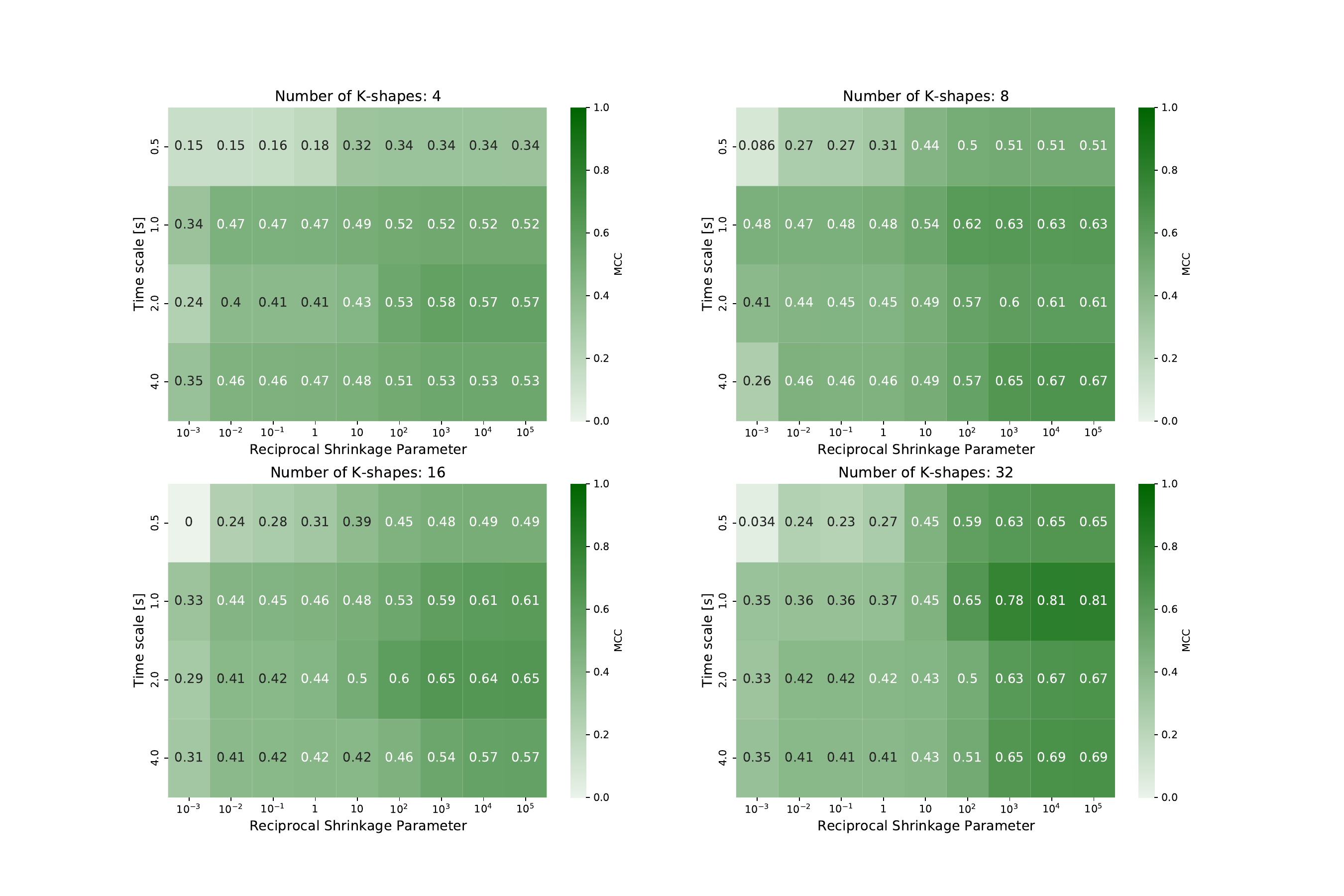}
    \caption{Sensitivity analysis for \gls{af}. Metric \gls{mcc}}
    \label{fig:sens_anal_af_mcc}
\end{figure}

\begin{figure}
    \centering
    \includegraphics[width= .9\textwidth]{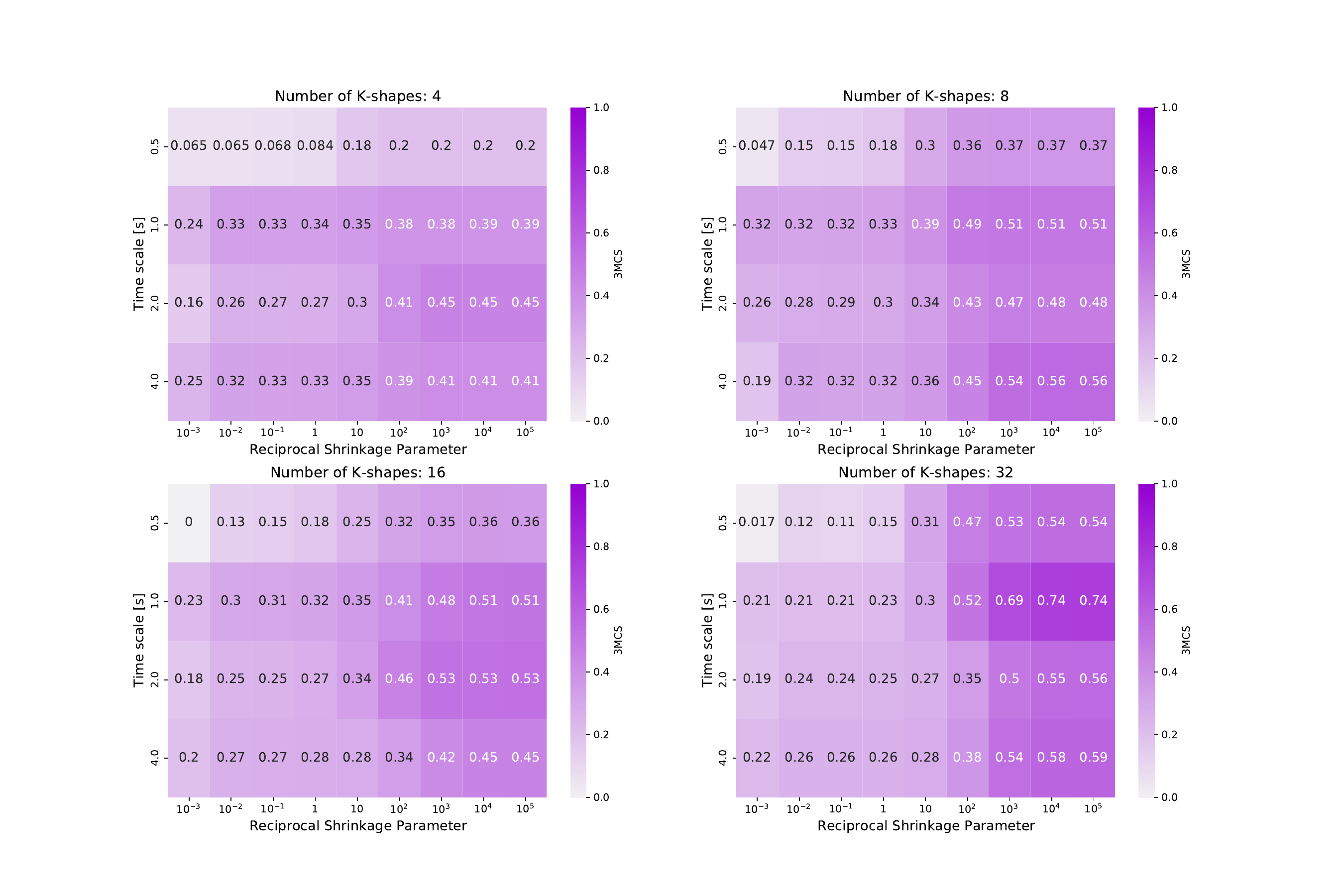}
    \caption{Sensitivity analysis for \gls{af}. Metric \gls{3mcs}}
    \label{fig:sens_anal_af_3mcs}
\end{figure}

\begin{figure}
    \centering
    \includegraphics[width= .9\textwidth]{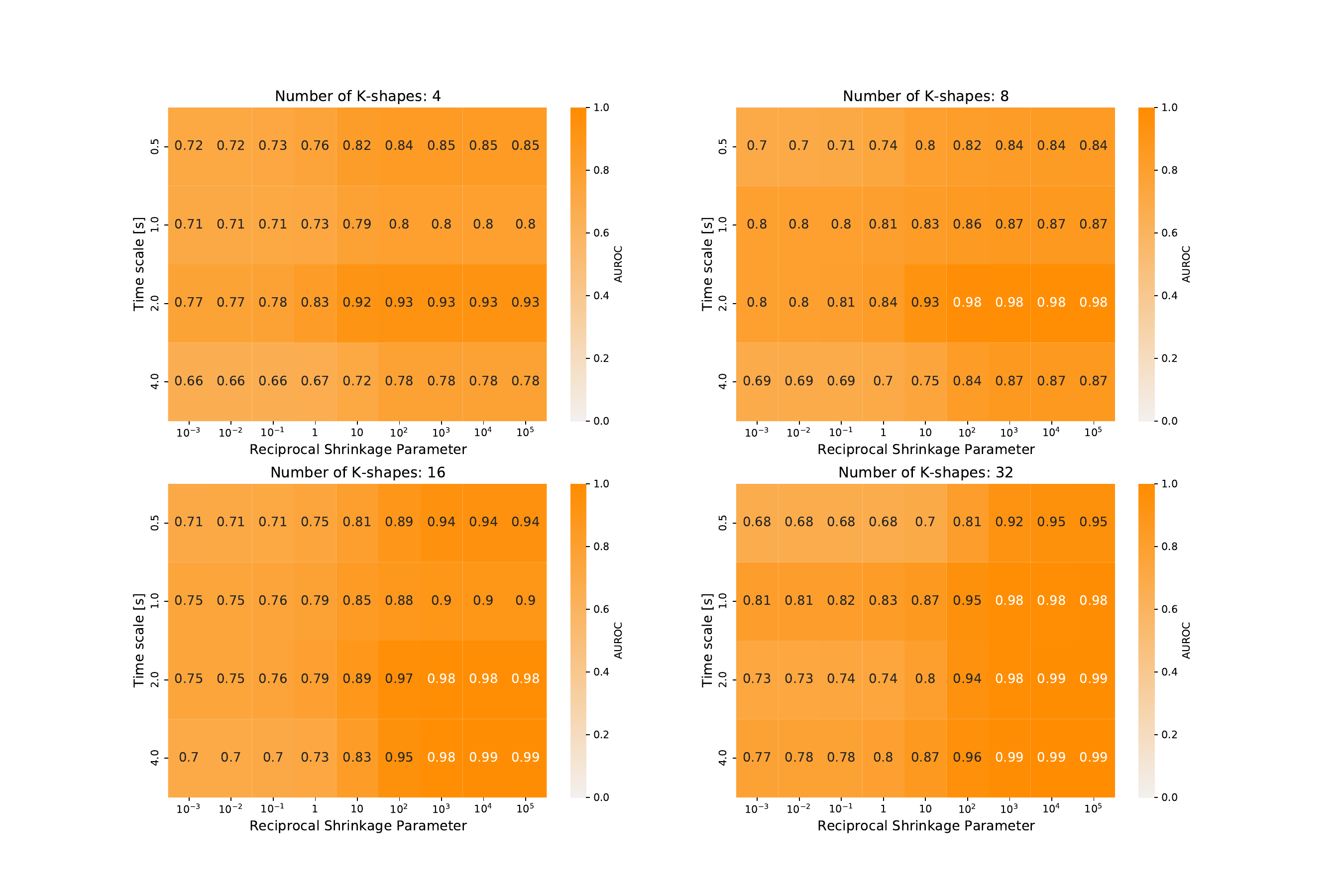}
    \caption{Sensitivity analysis for \gls{mi}. Metric \gls{auroc}}
    \label{fig:sens_anal_mi_auroc}
\end{figure}

\begin{figure}
    \centering
    \includegraphics[width= .9\textwidth]{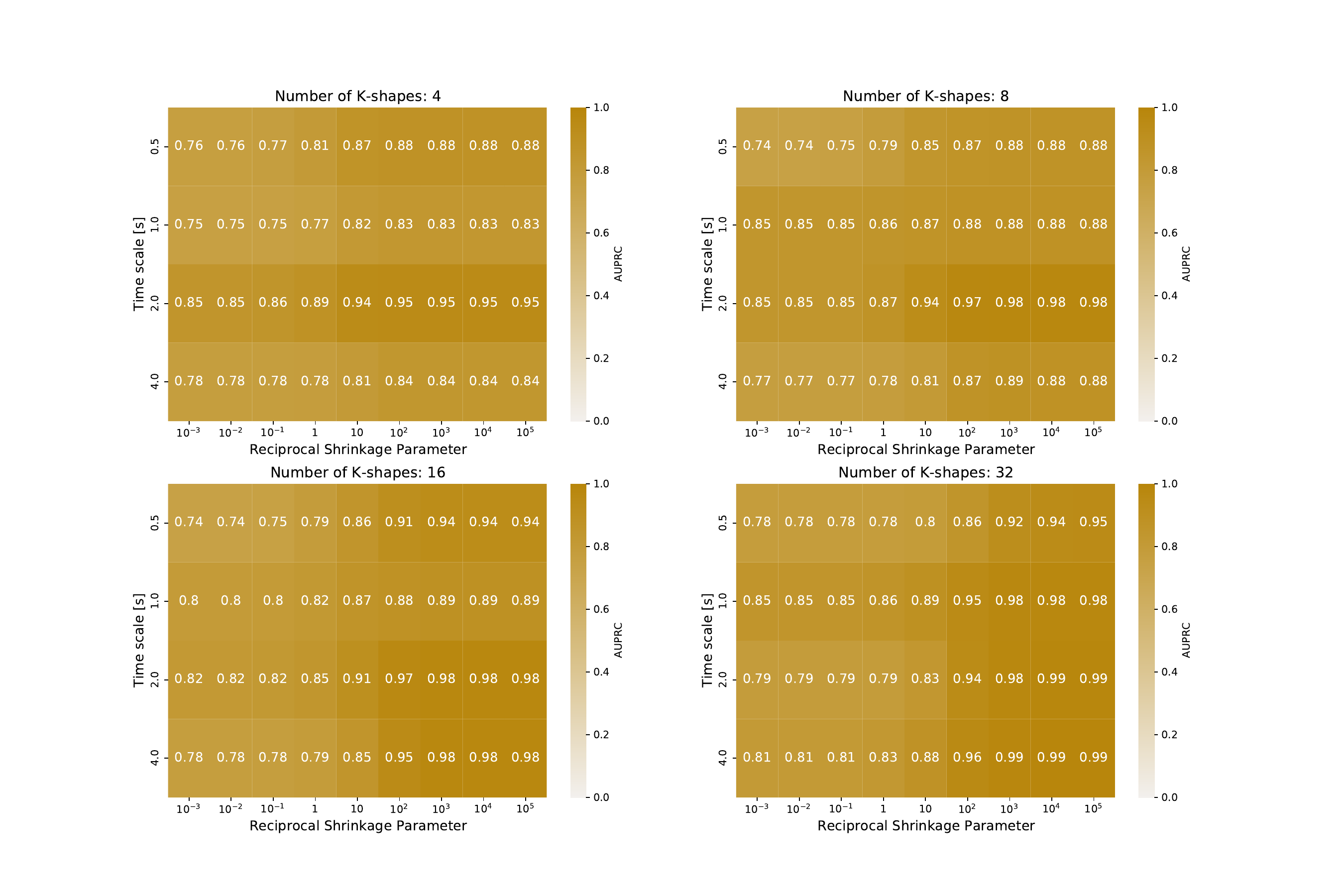}
    \caption{Sensitivity analysis for \gls{mi}. Metric \gls{auprc}}
    \label{fig:sens_anal_mi_auprc}
\end{figure}

\begin{figure}
    \centering
    \includegraphics[width= .9\textwidth]{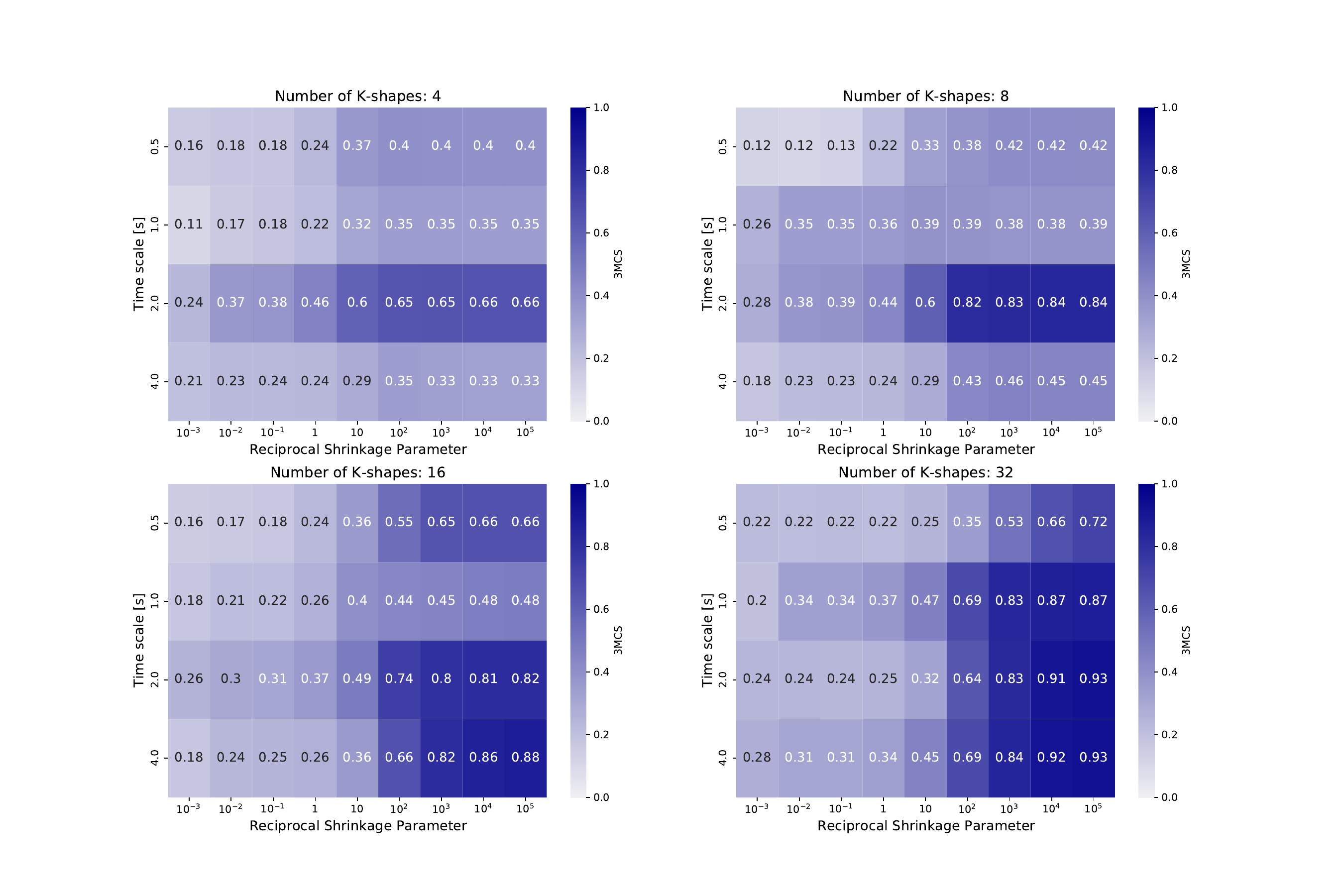}
    \caption{Sensitivity analysis for \gls{mi}. Metric \gls{3mcs}}
    \label{fig:sens_anal_mi_3mcs}
\end{figure}

\begin{figure}
    \centering
    \includegraphics[width= .9\textwidth]{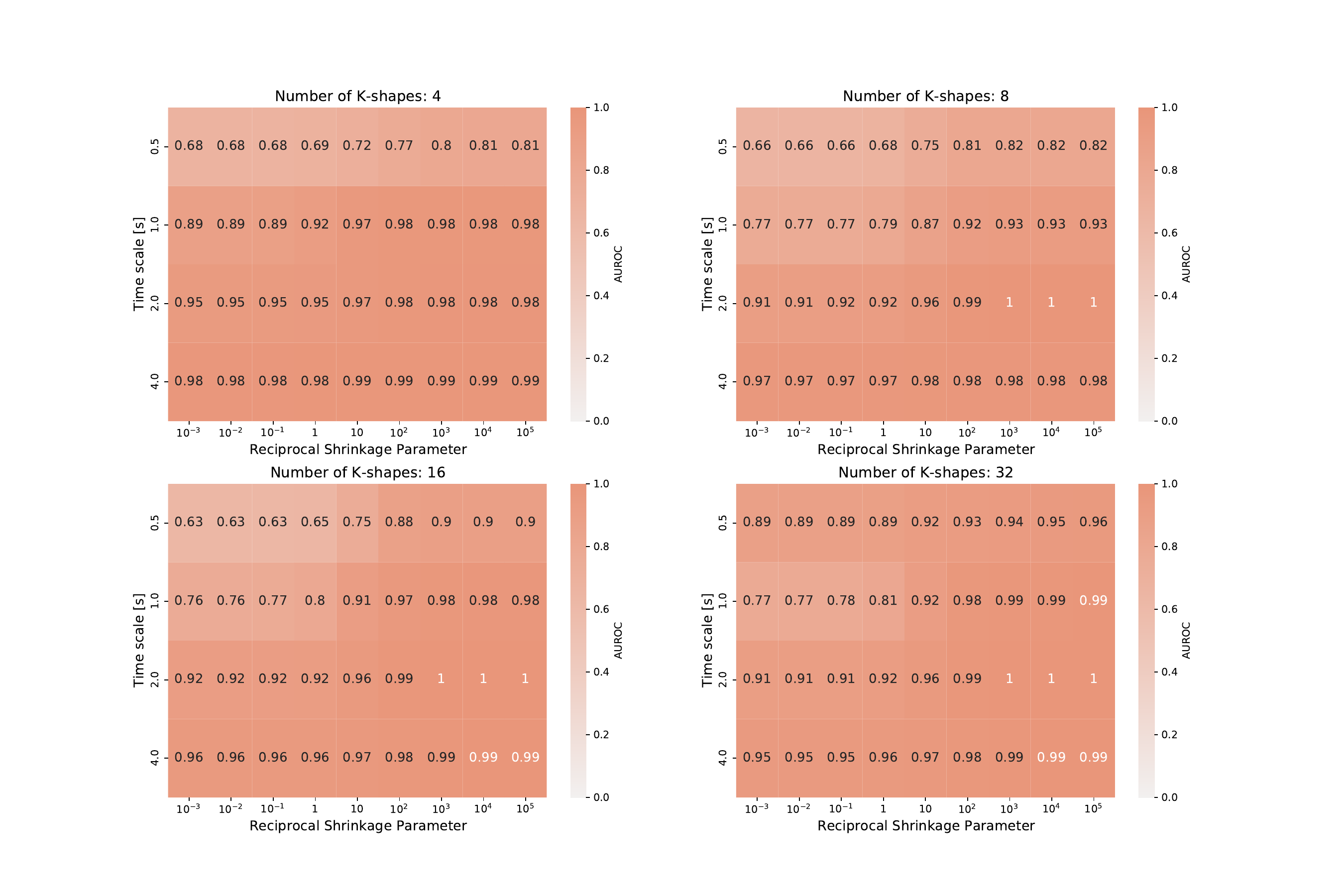}
    \caption{Sensitivity analysis for \gls{sbr}. Metric \gls{auroc}}
    \label{fig:sens_anal_sb_auroc}
\end{figure}

\begin{figure}
    \centering
    \includegraphics[width= .9\textwidth]{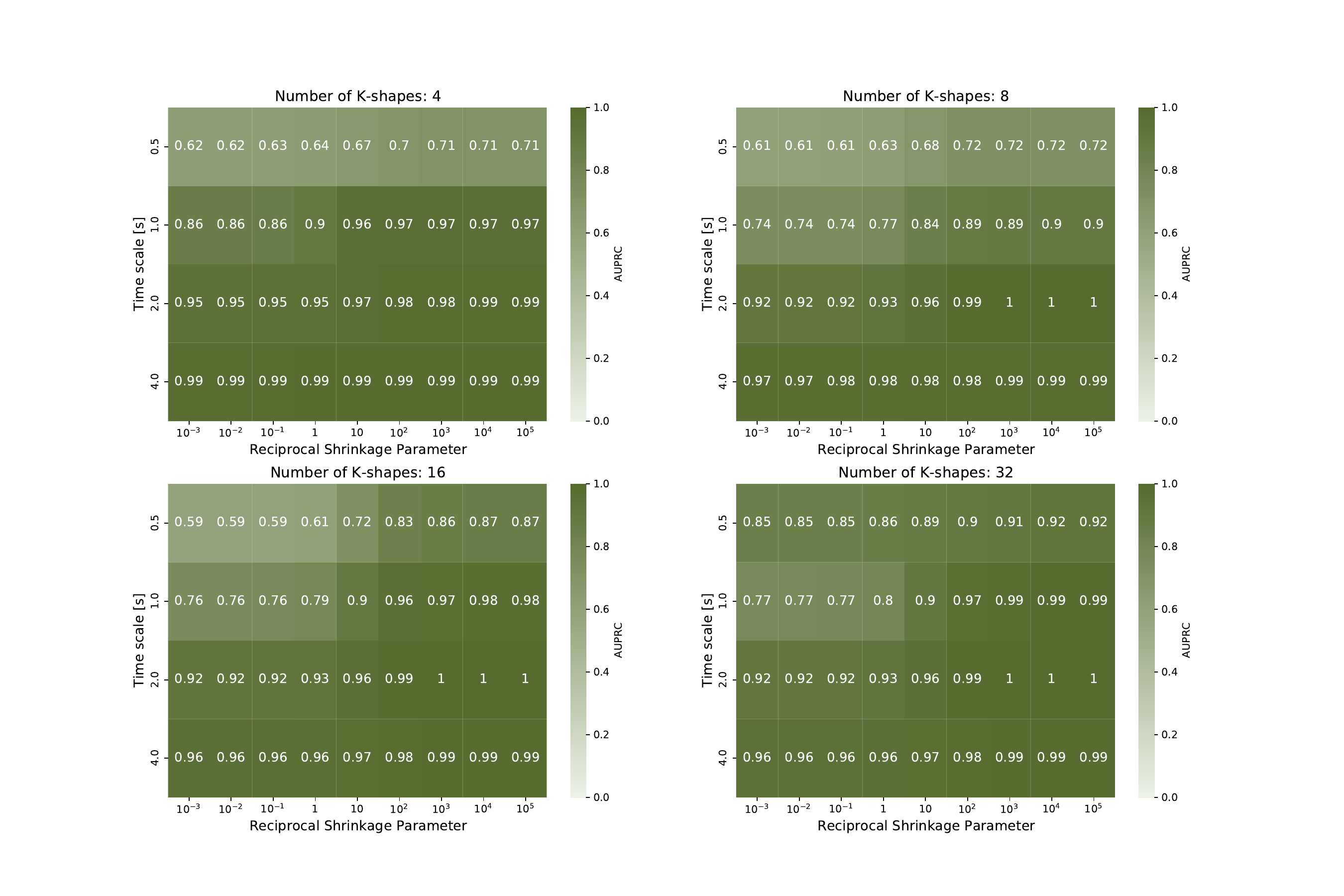}
    \caption{Sensitivity analysis for \gls{sbr}. Metric \gls{auprc}}
    \label{fig:sens_anal_sb_auprc}
\end{figure}

\begin{figure}
    \centering
    \includegraphics[width= .9\textwidth]{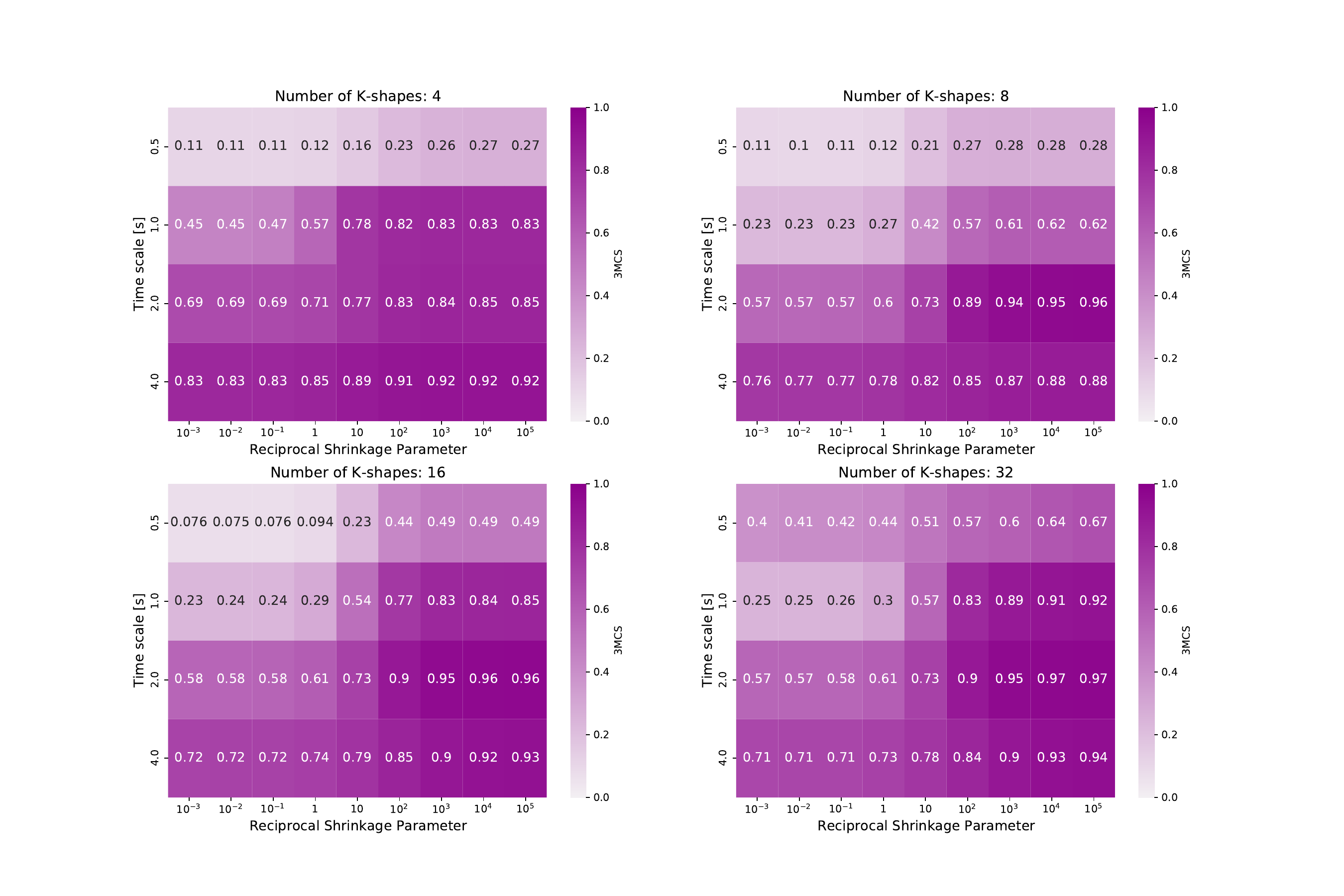}
    \caption{Sensitivity analysis for \gls{sbr}. Metric \gls{3mcs}}
    \label{fig:sens_anal_sb_3mcs}
\end{figure}

The feature extraction we proposed is then sensitive to both the amplitude of \gls{ks} centroids and the number of clusters (i.e., the number of  \gls{ks} centroids).
After pre-processing the extracted features with \gls{kpca}, we fed them into the \gls{lr} model. 
To determine the optimal feature extraction settings, we also considered the shrinkage parameter of the ridge regularization in the LR model. 
Consequently, we employed a grid-search method to explore various combinations of these three parameters.
For completeness, we evaluated each configuration by reporting the performance of the \gls{lr} model using four metrics: \gls{auroc}, \gls{auprc}, \gls{mcc}, and \gls{3mcs}, for each classification task.
For the \gls{af}; results are shown in Figures \ref{fig:sens_anal_af_auroc}-\ref{fig:sens_anal_af_3mcs}.
For the \gls{mi}; results are shown in Figures \ref{fig:sens_anal_mi_auroc}-\ref{fig:sens_anal_mi_3mcs}.
For the \gls{sbr}; results are shown in Figures \ref{fig:sens_anal_sb_auroc}-\ref{fig:sens_anal_sb_3mcs}.


\begin{figure}
    \centering
    \includegraphics[width= .9\textwidth]{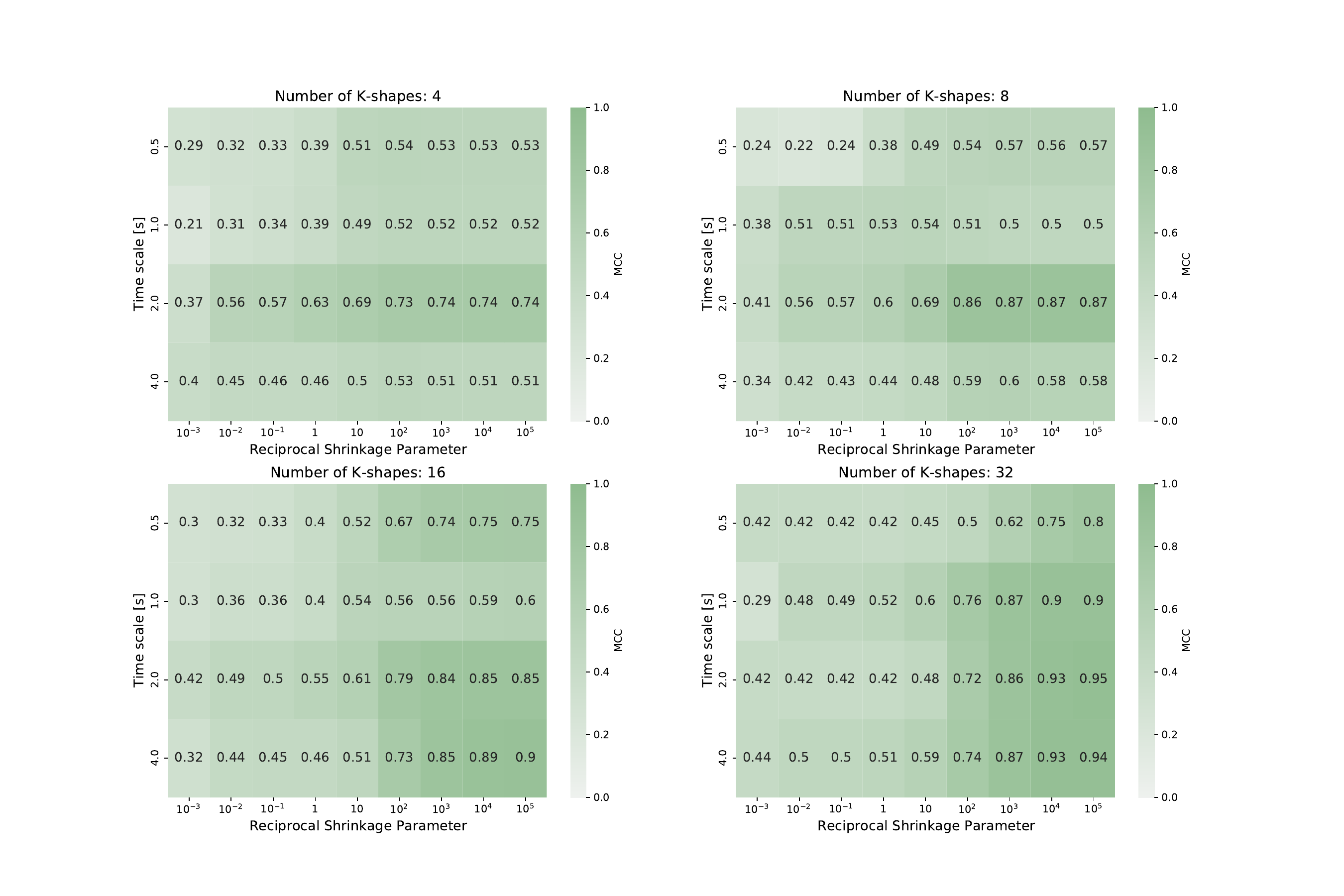}
    \caption{Sensitivity analysis for \gls{mi}. Metric \gls{mcc}}
    \label{fig:sens_anal_mi_mcc}
\end{figure}

\begin{figure}
    \centering
    \includegraphics[width= .9\textwidth]{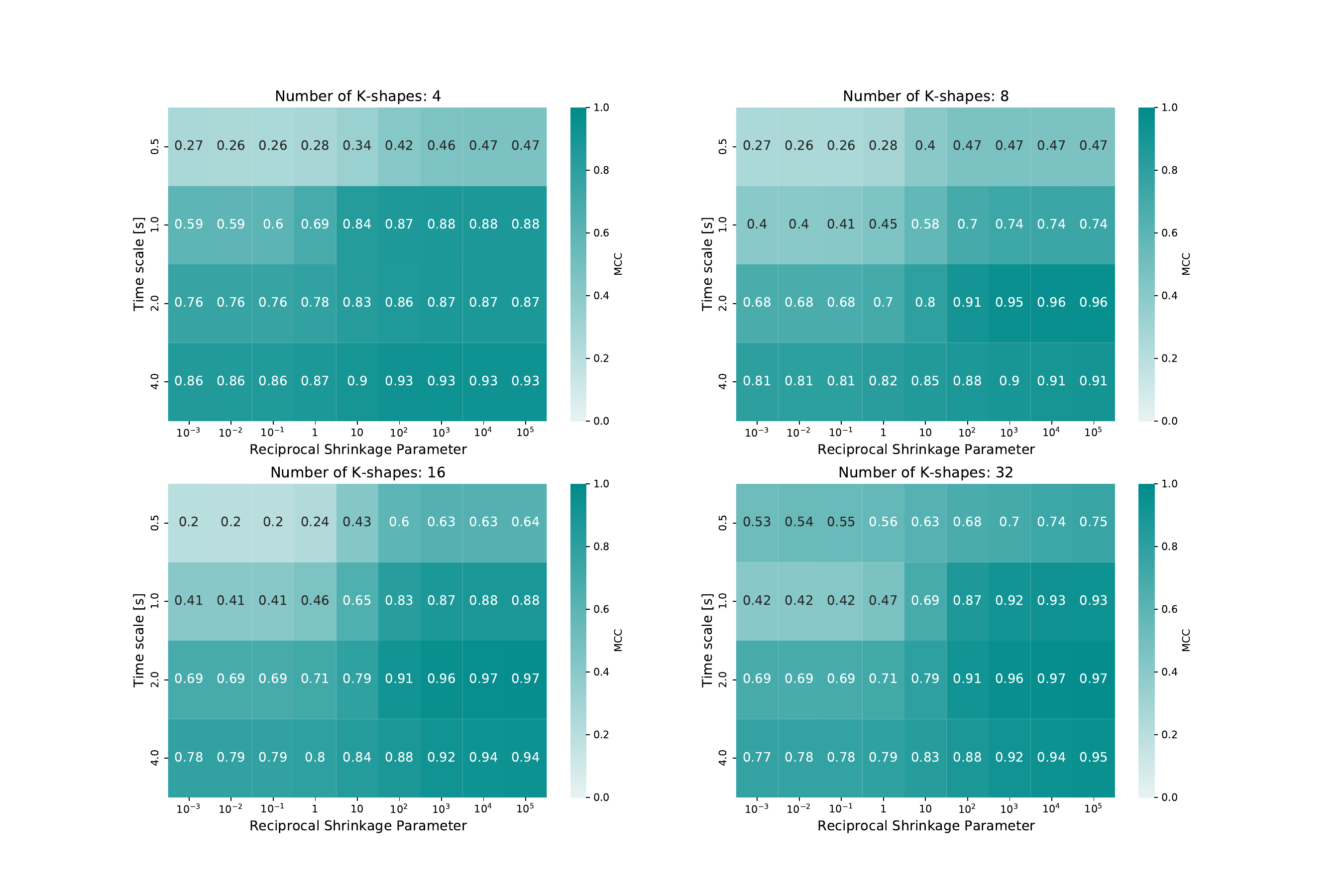}
    \caption{Sensitivity analysis for \gls{sbr}. Metric \gls{mcc}}
    \label{fig:sens_anal_mi_mcc}
\end{figure}

\section{Correlation Analysis Between Presence Features and KPCA Features}\label{apx: correlation_presence_KPCA_features}
As previously discussed, the feature separability achieved through \gls{kpca} enhances the predictive power of the \gls{lr} model. However, this separability can complicate the interpretation of \gls{lr}'s feature inspection in terms of \gls{ks} centroids. Since \gls{kpca} applies a non-linear transformation to the \gls{ks} centroids, we investigated which \gls{ks} centroids most significantly support the \gls{lr} predictions by examining the correlation between the presence of these centroids and the most important features in the \gls{lr} model.

\begin{figure}
    \centering
    \includegraphics[width= .5\textwidth]{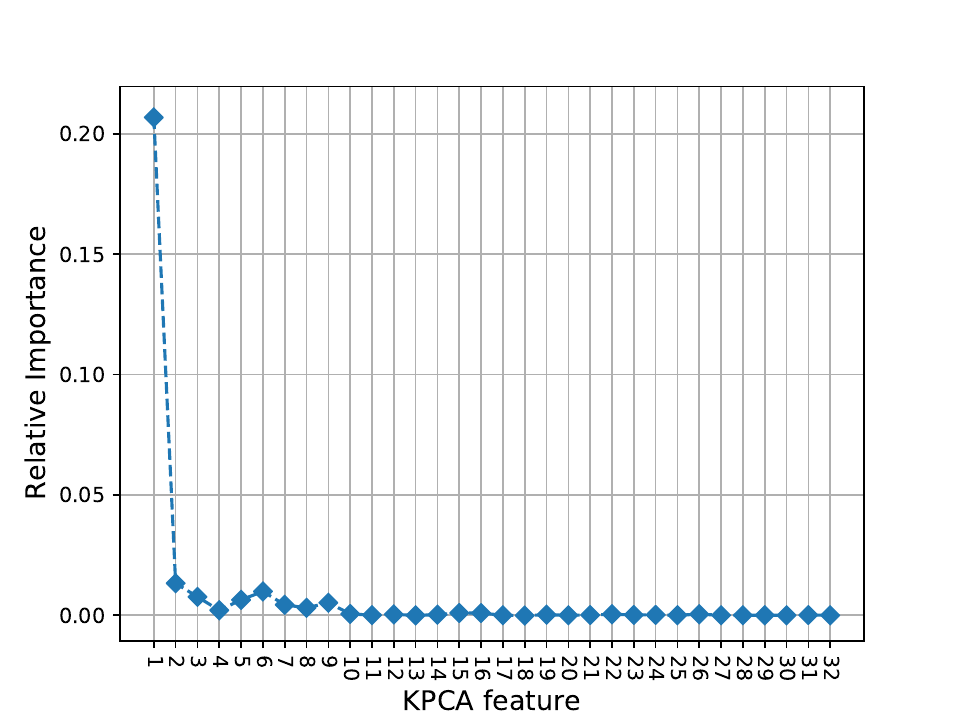}
    \caption{Feature inspection of \gls{lr} for \gls{af} classification}
    \label{fig: inspection_af}
\end{figure}

\begin{figure}
    \centering
    \includegraphics[width= .5\textwidth]{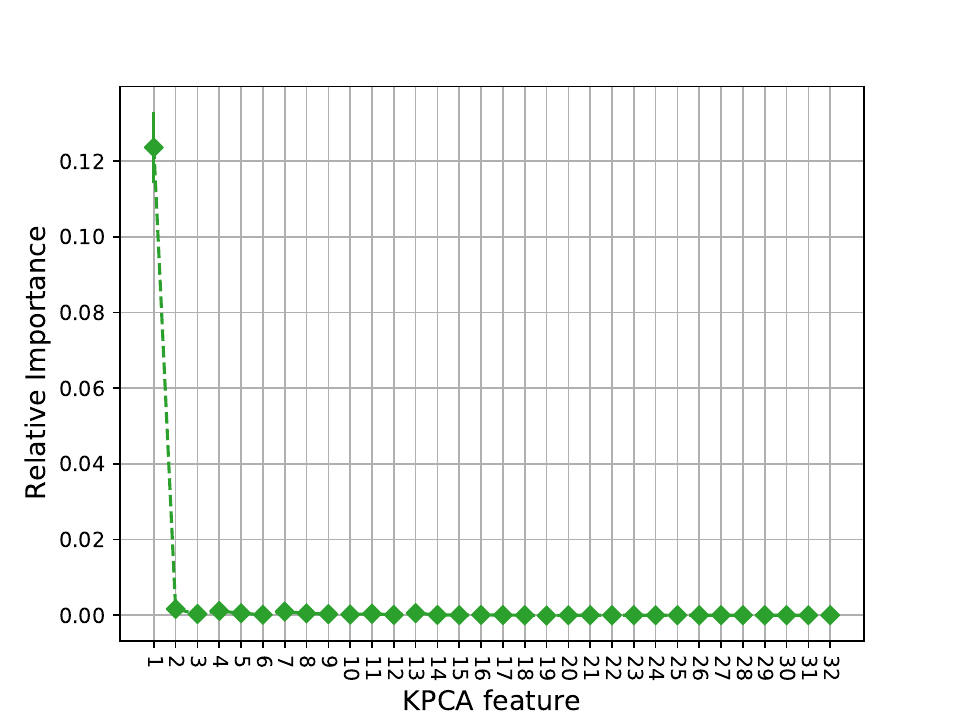}
    \caption{Feature inspection of \gls{lr} for \gls{mi} classification}
    \label{fig: inspection_mi}
\end{figure}

\begin{figure}
    \centering
    \includegraphics[width= .5\textwidth]{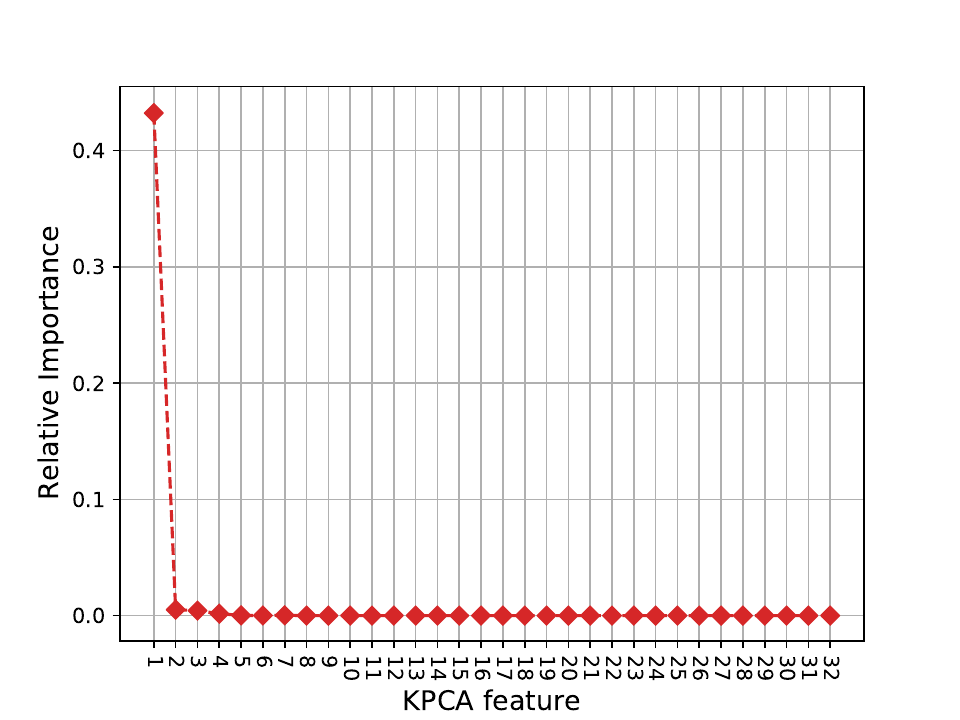}
    \caption{Feature inspection of \gls{lr} for \gls{sbr} classification}
    \label{fig: inspection_sb}
\end{figure}

The inspection of the \gls{lr} models' importance revealed that only one feature supports the predictions.
Such a results was confirmed for \gls{af}, \gls{mi}, and \gls{sbr}; see Figures \ref{fig: inspection_af}-\ref{fig: inspection_sb}

\begin{figure}
    \centering
    \includegraphics[width= .5\textwidth]{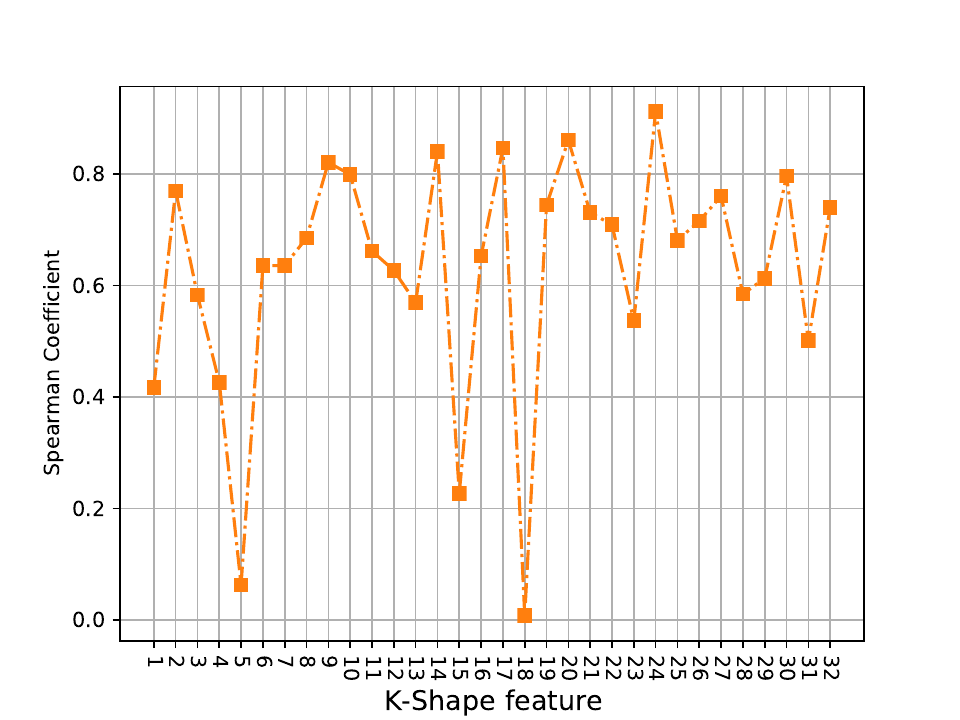}
    \caption{Connection between the most important feature of \gls{lr} and the \gls{ks} centroids. \gls{af} classification problem}
    \label{fig: connection_af}
\end{figure}

\begin{figure}
    \centering
    \includegraphics[width= .5\textwidth]{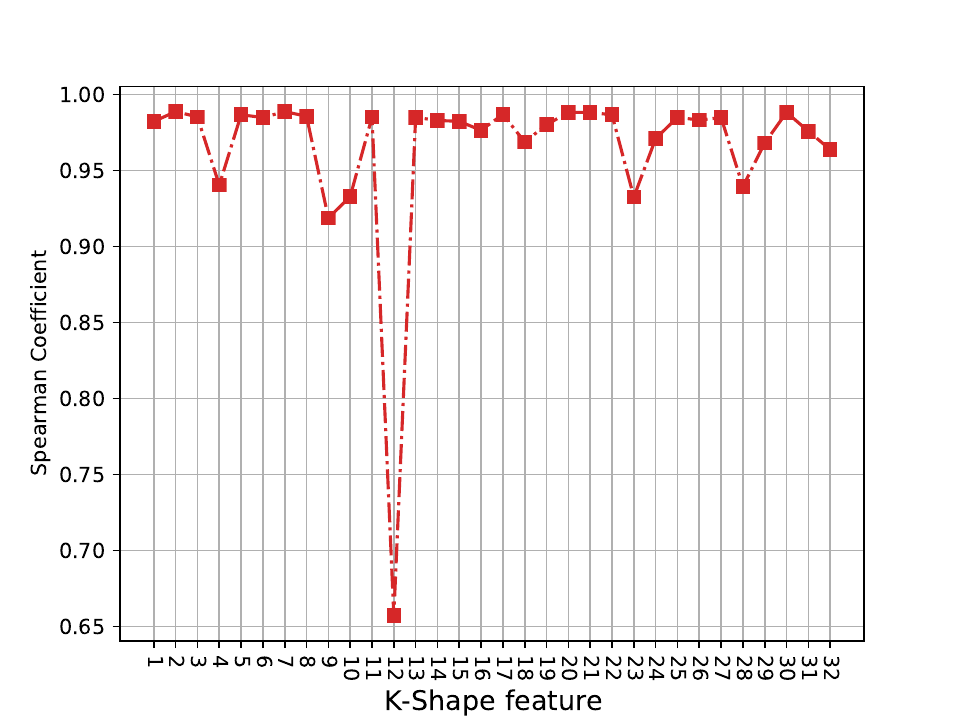}
    \caption{Connection between the most important feature of \gls{mi} and the \gls{ks} centroids. \gls{mi} classification problem}
    \label{fig: connection_mi}
\end{figure}

\begin{figure}
    \centering
    \includegraphics[width= .5\textwidth]{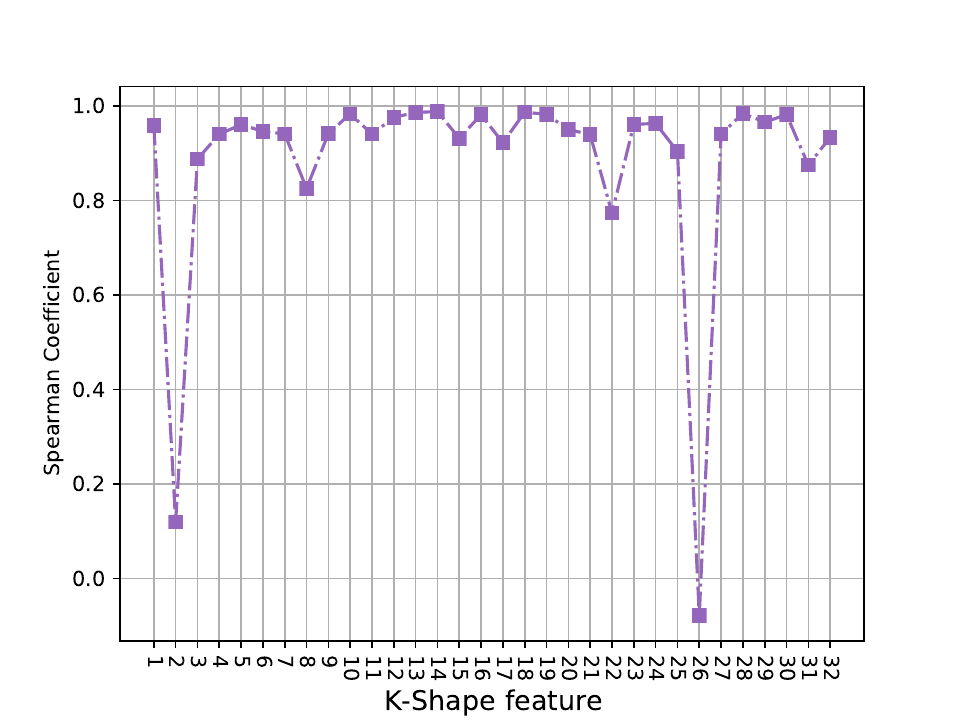}
    \caption{Connection between the most important feature of \gls{lr} and the \gls{ks} centroids. \gls{sbr} classification problem}
    \label{fig: connection_sb}
\end{figure}

When considering the connection between that feature and the \gls{ks} centroids, the Spearman test revealed plenty of gls{ks} centroids connected to that feature; see Figures \ref{fig: connection_af}-\ref{fig: connection_sb}.

\section{The Likelihood Ratio Test}\label{apx: Likelihood Ratio Test}

In this section, we shall introduce how the \gls{lrt} was readapted for the analysis of Section \ref{sec: Comapring_ Interpretable_Predictions}.
We consider two sets of regressors, one corresponding to the model that uses only \gls{ecg}-derived features, and the other for the full model incorporating both \gls{ecg}-derived features and baseline covariates (such as age and sex). 
Let the set of possible regressors for the \gls{ecg}-derived features model be defined as follows:

$$
W_0 = \left\{ \mathbf{w} = (w_1, \dots, w_n) \mid w_i \in \mathbb{R}, \, \forall i = 1, \dots, n, \, \|\mathbf{w}\| \leq M \right\}.
$$
where \( M \) is any real positive quantity. For the full model, which includes the additional covariates age and sex, the set of regressors is given by:
$$
W_1 = \left\{ \mathbf{w} = (w_1, \dots, w_n, w_{age}, w_{sex}) \mid w_i \in \mathbb{R}, \, \forall i = 1, \dots, n, \, w_{age} \in \mathbb{R}, \, w_{sex} \in \mathbb{R}, \, \|\mathbf{w}\| \leq M \right\}.
$$

We test the null hypothesis \( H_0 \), which assumes that the regressor vector \( \tilde{w} \) necessary to explain the complete \gls{ecg} dataset lies within \( W_0 \), i.e.,

$$
H_0 : \tilde{w} \in W_0.
$$
The alternative hypothesis is:
$$
H_1 : \tilde{w} \notin W_0.
$$

In this framework, the \gls{lrt} statistic is defined as:
\begin{equation}\label{eq: LRT_statistic}
\lambda_{LR} = 2 \log{\left( \frac{\sup_{\tilde{w} \in W_0}{\mathcal{L}(\tilde{w})}}{\sup_{\tilde{w} \in W_1}{\mathcal{L}(\tilde{w})}} \right)};    
\end{equation}

where $\mathcal{L}(\cdot) $ denotes the likelihood function. The value of \( \lambda_{LR} \) is expected to follow a chi-squared distribution with 2 degrees of freedom, as the difference in dimensionality between the two models is 2 (due to the inclusion of \( w_{age} \) and \( w_{sex} \)).

The \emph{p-value} of the test is given by:
$$
\mathbf{P}(\chi^{2}_{2} > \lambda_{LR}).
$$
This p-value determines whether we reject the null hypothesis, thereby assessing the significance of the baseline covariates in addition to the \gls{ecg}-derived features in explaining the data.

The \gls{lrt} was utilized during the validation phase of the full model. 
As stated in Section \ref{sec: Comapring_ Interpretable_Predictions}, the same folds used for validating the \gls{ecg}-derived features model were employed. This ensures a consistent evaluation of \eqref{eq: LRT_statistic}. Note that the evaluation of \eqref{eq: LRT_statistic} requires that only the training data be utilized.

We conducted a 5-fold cross-validation, which allowed us to formulate hypothesis tests corresponding to the number of folds involved. 
To account for the potential of Type I Errors, we applied the Bonferroni correction to rescale the significance level, setting it to $\alpha \to \alpha/5$. 
Thus, with a nominal significance level of $\alpha= 0.05$, the adjusted significance level becomes $0.01.$

\end{appendices}

\printglossary[type=acronym, title=List of Acronyms]

\section*{Acknowledgments}
The authors would like to acknowledge the resources and tools that facilitated this research.

\section*{Code availability}
The code developed in this study is available at 
\begin{center}
    \url{https://github.com/glancia93/Constructing-Interpretable-Prediction-Models-with-1D-DNNs}.
\end{center}

\section*{Conflicts of interest}
The authors declare no conflicts of interest.

\bibliographystyle{apalike} 
\bibliography{references}  

\end{document}